\documentclass[a4paper,fleqn,usenatbib,useAMS, twocolumn]{mnras}

\usepackage{graphicx}	
\usepackage{amsmath}	
\usepackage{amssymb}	
\usepackage{multicol}        
\usepackage{bm}		
\usepackage{pdflscape}	
\usepackage{listings}
\usepackage{alltt}
\usepackage{xcolor}
\usepackage{booktabs}
\usepackage{tablefootnote}

\newcommand{\Msun}{M_{\rm \odot}}

\title[Gas Cooling With An Exact Time Integration Scheme]{Gas Cooling in Hydrodynamic Simulations with An Exact Time Integration Scheme}

\author[ Q. Zhu et al.]
{Qirong Zhu$^{1,2}$\thanks{E-mail: qxz125@psu.edu}, 
Britton Smith$^{3}$, and
Lars Hernquist$^{2}$ 
\vspace{0.5cm}\\
\parbox{\textwidth}
{\small $^{1}$Department of Astronomy \& Astrophysics; Institute for Cosmology and Gravity, The Pennsylvania State University, PA 16802, USA\\
$^{2}$Harvard-Smithsonian Center for Astrophysics, Harvard University, 60 Garden Street, Cambridge, MA 02138, USA\\
$^{3}$San Diego Supercomputer Center, University of California, San Diego, 10100 Hopkins Drive, La Jolla, CA 92093, USA\\
}}

\begin{document}

\date{Accepted 2017 May 26. Received 2017 May 25; in original form 2017 April 21}

\pagerange{\pageref{firstpage}--\pageref{lastpage}} \pubyear{2016}

\maketitle

\label{firstpage}

\begin{abstract}
We implement and test the exact time integration method proposed by
\cite{Townsend2009} for gas cooling in cosmological hydrodynamic
simulations. The errors using this time integrator for the internal energy 
are limited by the resolution
of the cooling tables and are insensitive to the size of the
timestep, improving accuracy relative to explicit or
implicit schemes when the cooling time is short. We compare
results with different time integrators for gas
cooling in cosmological hydrodynamic simulations. We find that the
temperature of the gas in filaments before accreting into dark matter
halos to form stars, obtained with the exact cooling integration, lies
close to the equilibrium where radiative cooling
balances heating from the UV background.  For comparison, the gas
temperature without the exact integrator shows
substantial deviations from the equilibrium relation.  Galaxy stellar
masses with the exact cooling technique agree reasonably
well, but are systematically lower than the results
obtained by the other integration schemes, reducing the
need for feedback to suppress star formation.
Our implementation of the exact cooling technique is provided
and can be easily incorporated into any hydrodynamic code.
\end{abstract}


\section{Introduction}

\begin{figure*}
\begin{center}
\begin{tabular}{lll}
\includegraphics[trim=0.2cm 0.2cm 0.0 0.2cm,clip,width=0.33\linewidth]{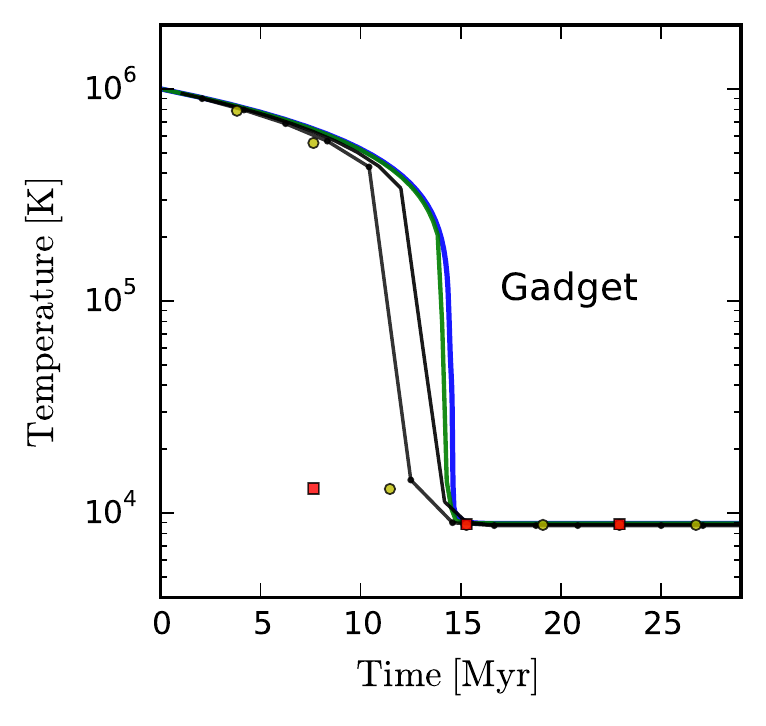}
\includegraphics[trim=0.2cm 0.2cm 0.0 0.2cm,clip,width=0.33\linewidth]{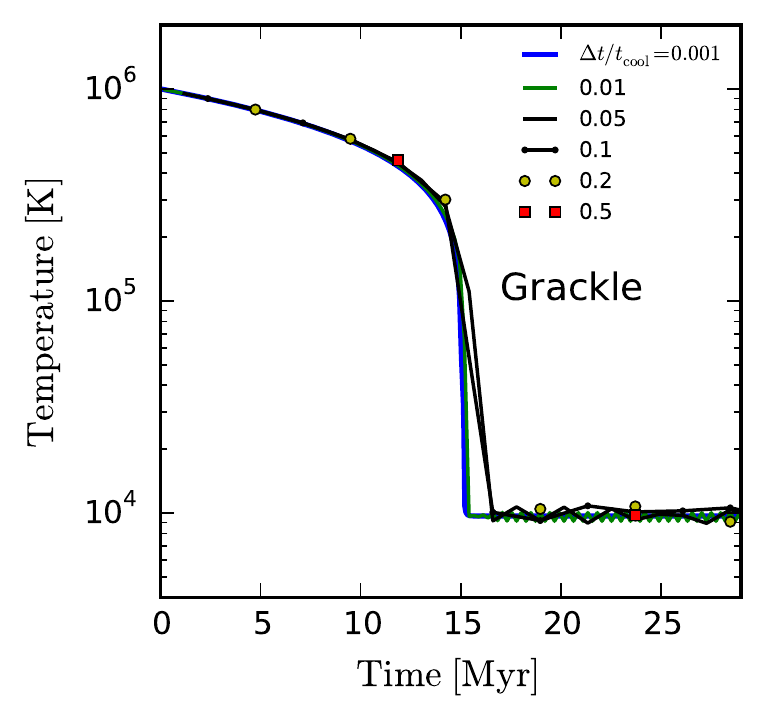}
\includegraphics[trim=0.2cm 0.2cm 0.0 0.2cm,clip,width=0.33\linewidth]{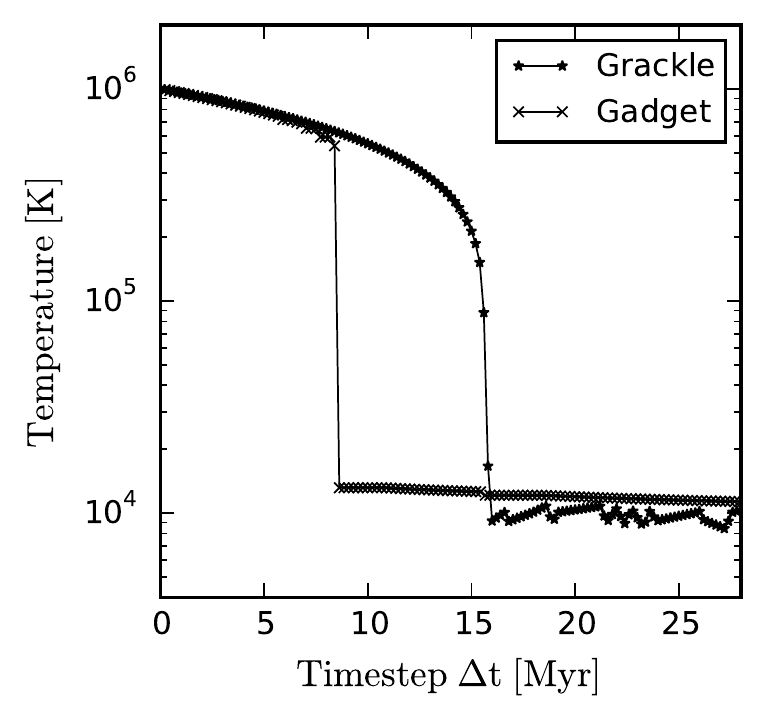}\\
\end{tabular}
\caption{\label{fig:cooling_cell_grackle} \textit{Left Panel}: Impact of timestep size 
on the predicted gas temperature in the constant density 
cooling test using the {\sc Gadget} cooling method.
The thick blue line shows the predicted gas temperature
started from $10^6$ K using a very fine timestep of $10^{-3}\, t_{\rm cool}$. The other lines are the
predicted temperature using different timestep sizes. Once the timestep is comparable to 
$t_{\rm cool}$, substantial errors are found in the predicted temperature. 
\textit{Middle Panel}: The constant cooling test using the cooling routine in {\sc Grackle}. 
Oscillations in gas temperature around $10^4$ K are present for all the curves with 
$\Delta t \ge 0.01\, t_{\rm cool}$, which highlights the stability issue of this method.
\textit{Right Panel}: Predicted temperature as a function of a \textit{single} timestep $\Delta t$.
Once the timestep size is larger than 8.5 Myr, gas temperature
with {\sc Gadget} has already dropped to $\sim 10^4$ K. The overall behavior of gas cooling
with {\sc Grackle} agrees well with the middle panel. However, the oscillations in temperature
are more irregular than that in the middle panel.}
\end{center}
\end{figure*}

Energy loss via radiative cooling in hot temperature plasmas plays a
crucial role in the formation and evolution of galaxies
\citep{Blumenthal1984}. Hydrodynamic simulations are routinely used to
study galaxy formation, including radiative cooling, star formation,
and other relevant processes.  Radiative cooling itself together with
gravity in cosmology represents a complex problem that cannot be
addressed analytically \citep[e.g.][]{Birnboim2003, Keres2005,
  Nelson2016}.

Recently, many studies have focused on the accuracies of hydrodynamic
solvers \citep{Arepo, Bauer2012, Gizmo, Zhu2015} and the differences
between different codes in galaxy formation studies (e.g.,
\citealt{Sijacki2012, Keres2012, Vogelsberger2012, Torrey2012,
Nelson2013, Zhu2016}). While state-of-the-art codes such as 
{\sc Arepo} and {\sc Gizmo} are second-order accurate in time (and
space), the cooling step is often only first-order.  It is thus also
timely to review the performance of the cooling method now being used in
current hydrodynamic simulations.

Typically, radiative cooling in cosmological hydrodynamic codes is
either handled with an explicit scheme
\begin{equation}
\\
u_i(t+\Delta t) = u_{i}(t) + \Delta t \cdot \Lambda(u_i(t))
\label{eq:explicit}
\end{equation}
or an implicit one
\begin{equation}
\\
u_i(t+\Delta t) = u_{i}(t) + \Delta t \cdot \Lambda(u_i(t+\Delta t)), 
\label{eq:implicit}
\end{equation}
depending on whether the starting time $t$ or the ending time $t+\Delta t$, 
with $\Delta t$ the timestep,
is used to evaluate the cooling rates. Equations~(\ref{eq:explicit}, \ref{eq:implicit})
operate on the thermal energy per unit mass $u_i$ for gas element $i$, a common thermodynamic
state used in numerical simulations.
Note that in these two Equations and the rest of the paper, we
have normalized the cooling/heating rate $\Lambda$ with a factor ${n_{\rm{H}, i}^2 / \rho_{i}}$ 
to be consistent with the units in the cooling table. 

The explicit scheme is straightforward to implement. However,
stability requires that the timestep be controlled carefully.  The
integration of the specific internal energy in {\sc Grackle} \citep{Smith2016}
is close to a first-order explicit method. The implicit method for
radiative cooling is common among SPH codes \citep{Hernquist1989}, in
particular {\sc Gadget }\citep{Springel2001}, as well as 
in AMR codes such as {\sc RAMSES}
\citep{RAMSES}. While this approach is stable, it entails inaccuracies
due to the finite timestep and errors arising from multiple zero
points for certain rooting finding methods \citep{Townsend2009}.

To control the accuracy or to maintain the stability of explicit
techniques, the internal energy update in each sub-cycle is subject
to empirical constraints such as the ``10 percent rule''
\citep{Teyssier2015, Smith2016}. For explicit schemes, sub-cycling is
mandatory due to its unstable behavior with respect to the timestep.
However, substantial errors result once the timestep is comparable
to the cooling time $t_{\rm cool}$ even when sub-cycling is
applied. This can be seen in Figure~\ref{fig:cooling_cell_grackle},
which shows the evolution of the temperature in the constant density
cooling test of \cite{Smith2016} using a density of $0.1\,\rm{cm}^{-3}$. 
We use the cooling routines
available in {\sc Grackle} and {\sc Gadget}, respectively. We vary the
timestep to see how the temperature of the gas evolves from $10^6\,
\rm{K}$ as a function of time. The cooling rate is calculated for
metal-free gas exposed to the UV background of
{\protect\cite{Haardt2012}} at redshift $z = 0$. For a timestep
comparable to $t_{\rm cool}$, errors appear in the predicted
temperature using the {\sc Gadget} cooling method. The
accuracy is greatly
improved with {\sc Grackle} due to sub-cycling. Nevertheless,
fluctuations in the gas temperature around $10^4\, \rm{K}$ are present,
which is still noticeable for $\Delta t=0.01\, t_{\rm cool}$\footnote{
  In the case of net heating, we find that the impact of timestep size
  on the evolution of the gas temperature is also present for both
  implicit and explicit schemes.}.

We also repeat a test in \cite{Townsend2009} (their Figure 1 and 2), 
which is shown in the right panel with
the predicted gas temperature as a function of a \textit{single} timestep. 
This test represents the situation where the timestep size can span a large range 
in hydrodynamics simulations. The implicit method cools the gas too quickly once
the timestep size is larger than 8.5 Myr. While the gas temperature is overall stable, 
it is slightly above the result shown in the left panel. Thanks to sub-cycling, 
gas temperature with {\sc Grackle} agrees well with the result with very fine timesteps. 
However, the oscillations in gas temperature appear more irregular than that in the 
middle panel.

For the explicit and implicit methods, the evolution of internal
energy (gas temperature) is treated as an ODE system. A new
integration scheme proposed by \cite{Townsend2009} is based on the
following observation. For gas cooling, unlike a usual ODE system, we
already have the \textit{full knowledge} of all the future
evolution, which is simply expressed by the cooling curves.  Thus, one
can avoid the barrier presented by the stiff source terms and
analytically or numerically integrate the change of internal energy
along the cooling curves instead. Using this approach, under the
assumption that the cooling curves depend on the temperature data
points (power law or linear), one can simply and uniquely map an array
of internal energies to a time array.

In this paper, we implement and extend the method of
\cite{Townsend2009} to gas cooling in cosmological hydrodynamic
simulations, including cooling functions for gas exposed to a UV
background. We describe a set of simple algebraic equations to be
solved in Section~\ref{sec:method}. In Section~\ref{sec:results}, we
test the validity of the algorithm and confirm that the accuracy of
the exact time integration does not rely on the timestep size. We then
apply this cooling method to a cosmological hydrodynamic simulation
with gas cooling and star formation, and compare the results using the
{\sc Gadget} cooling method and {\sc Grackle}. We discuss our
findings in Section~\ref{sec:discussion} and provide conclusions
in Section~\ref{sec:conclusions}.
Our implementation
of the exact cooling technique is publicly available. 

\section{Methods}
\label{sec:method}

We discuss an implementation of the exact integration scheme with
the cooling/heating curves computed with the
photoionization code {\sc CLOUDY} \citep{Cloudy}.  A
redshift-dependent UV ionizing background is supplied to {\sc cloudy}
to model the heating rate due to young stars and AGNs. The essential
part of ``exact-integration'' is to integrate the possibly stiff
cooling/heating source term numerically by assuming that these rates
vary linearly\footnote{For the assumption of power-law dependence, we
refer the reader to \cite{Townsend2009}.}  between the temperature
grid points. Thus, numerical errors associated with this integration
scheme entirely arise from the above assumption, so that they are of
the same order as the linear interpolation errors. Such an
approach, compared with the popular implicit time
integration method, shows greatly improved accuracy. In light of the recent
developments of hydrodynamic solvers in computational galaxy formation
and evolution studies, we here consider whether or not errors due
to radiative cooling are important remaining sources of
error that should be addressed when designing progressively more
sophisticated models of star formation and feedback.

For each particle or cell $i$, the approach first locates its
position in redshift-density grids $[z, \log(n_{\rm H})]$ and then
bi-linearly interpolates the cooling / heating rates along all the
temperature data points ranging from $[10, 10^9]\, \rm{K}$. Note that
contributions for both primordial gas composition (H+He) and 
metal-enriched gas can be calculated. In our implementation, we have combined 
the mean molecular weight $\mu$ and gas temperature $T$ into the specific
internal energy $u$ through
\begin{equation}\\
u = \frac{kT}{(\gamma-1) \mu m_{\rm H}}, 
\label{eq:temp_u_conversion}
\end{equation}
such that the interpolations are carried out between the specific internal energy $u$ and  
\begin{equation}
\\
Y(u) = \frac{\Lambda_{\rm ref}}{u_{\rm ref}} \int_{u}^{u_{\rm ref}} \frac {{\rm d}u} {\Lambda(u)},
\end{equation}
a dimensionless temporal evolution function introduced by 
\cite{Townsend2009} in their Equation (24). 
This function measures the total time it takes to cool the gas 
normalized by a reference cooling time.

We denote the upper internal energy in the cooling table as $u_{\rm
  ref}$ and the cooling rate at that temperature as $\Lambda_{\rm
  ref}$. We use $u_{\rm ref}$ at $10^9\, \rm{K}$ for the adopted
cooling table, but it can be chosen at any temperature
higher than the current state. The computation of $Y(u)$ can be
performed in a piece-wise linear fashion on all the grid data
points. In each segment $[u_1, u_2]$, we assume the cooling rate
varies linearly from $\Lambda_1$ to $\Lambda_2$ as
\begin{equation}
\\
\Lambda(u) = \frac{\Lambda_2 - \Lambda_1}{u_2 - u_1}(u - u_1) + \Lambda_1, \text{if} \ \Lambda_2 \ne \Lambda_1.
\end{equation}
The change in
$\delta Y$ in such a segment can be analytically calculated as
\begin{equation}
\\
\delta Y = \frac{\Lambda_{\rm ref}}{u_{\rm ref}} A \log (\frac{B+u_2}{B+u_1}), \text{if} \ \Lambda_2 \ne \Lambda_1, 
\label{eq:deltay1}
\end{equation}
or 
\begin{equation}
\\
\delta Y = \frac{\Lambda_{\rm ref}}{u_{\rm ref}} \frac{u_2-u_1}{\Lambda_1}, \text{if} \ \Lambda_2 = \Lambda_1.
\label{eq:deltay2}
\end{equation}
In Equation~(\ref{eq:deltay1}) or Equation~(\ref{eq:deltay2}), we have defined $A$ and $B$ as 
\begin{equation}
\\ A \equiv  \frac{u_2 - u_1}{\Lambda_2 - \Lambda_1}, \\
B \equiv \frac{u_2\Lambda_1 - u_1\Lambda_2}{\Lambda_2 - \Lambda_1}.
\end{equation}
A monatomic curve of $Y(u)$ can be now summed up cumulatively using 
\begin{equation}
\\ Y(u) = \sum^{u}_{u_{\rm ref}}\delta Y, 
\label{eq:y}
\end{equation}
which starts from a reference internal energy $u_{\rm ref}$ to current internal energy $u$. 

To compute the internal energy for gas element $i$ at the next step $t+\Delta t$ , we find the expected change in $Y(u_i(t))$ as: 
\begin{equation}
\\
\Delta Y =	\frac{u_{i}(t)} {\Lambda_{u_{i}}(t)}\frac{\Lambda_{u_{\rm ref}}}{u_{\rm ref}}\frac{\Delta t}{t_{\rm cool}}
\label{eq:deltay3}
\end{equation}

Following Equation~(\ref{eq:deltay3}), we identify the segment in $Y$ that contains  
$Y(u_i)+\Delta Y$. The final step is to  analytically invert Equation~(\ref{eq:deltay1}) or
Equation~(\ref{eq:deltay2}) to find the predicted internal energy 
at the next time step. The above approach can be trivially extended to a net heating regime. 
One just starts from the lowest temperature as $u_{\rm ref}$ in the cooling table to compute 
$Y(u)$ and other quantities.

\section{Results}

The above equations are straightforward to implement and we provide an example C 
program\footnote{\url{https://goo.gl/UqKv68}}, which can be incorporated into codes 
such as {\sc Gadget} or {\sc Gizmo}. 

\label{sec:results} 

\begin{figure}
\begin{center}
\begin{tabular}{c}
\vspace{-0.2cm}
\includegraphics[width=\linewidth]{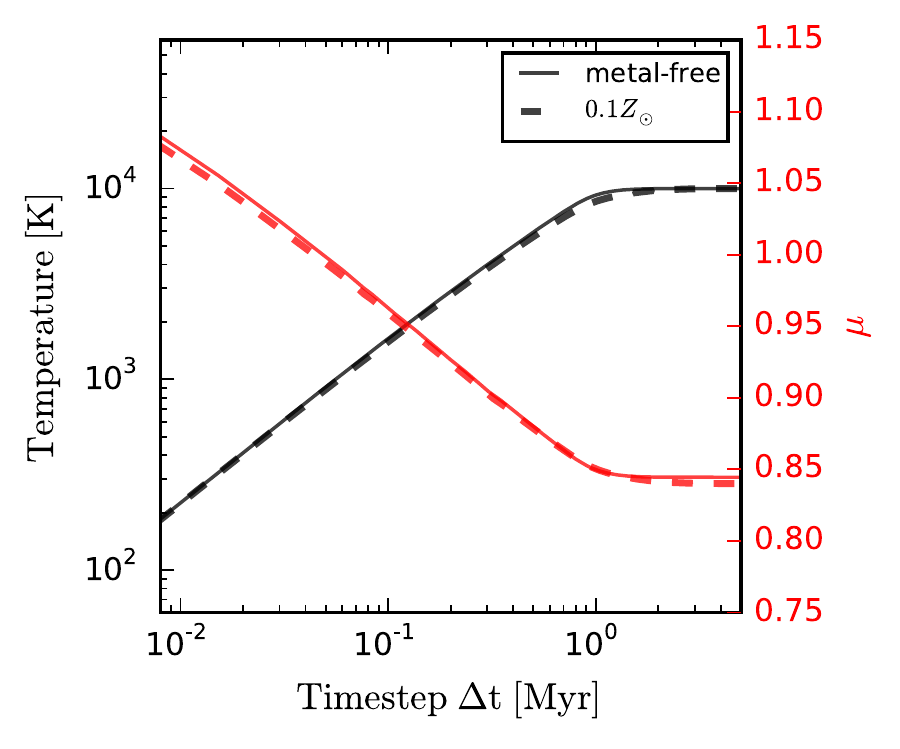}\\
\vspace{-0.2cm}
\includegraphics[width=\linewidth]{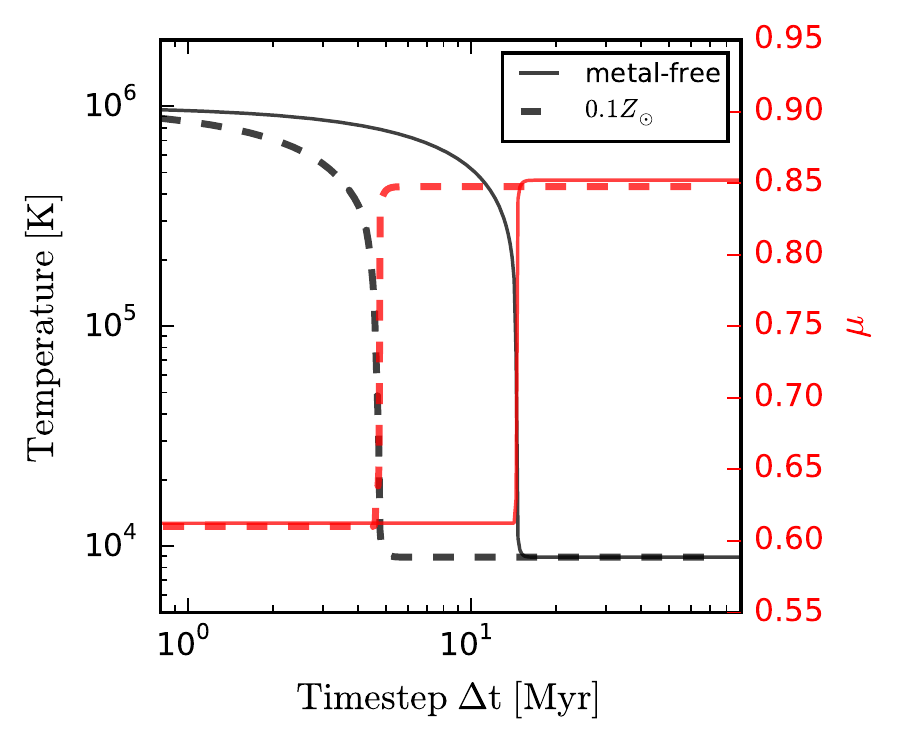}\\
\vspace{-0.2cm}
\includegraphics[width=\linewidth]{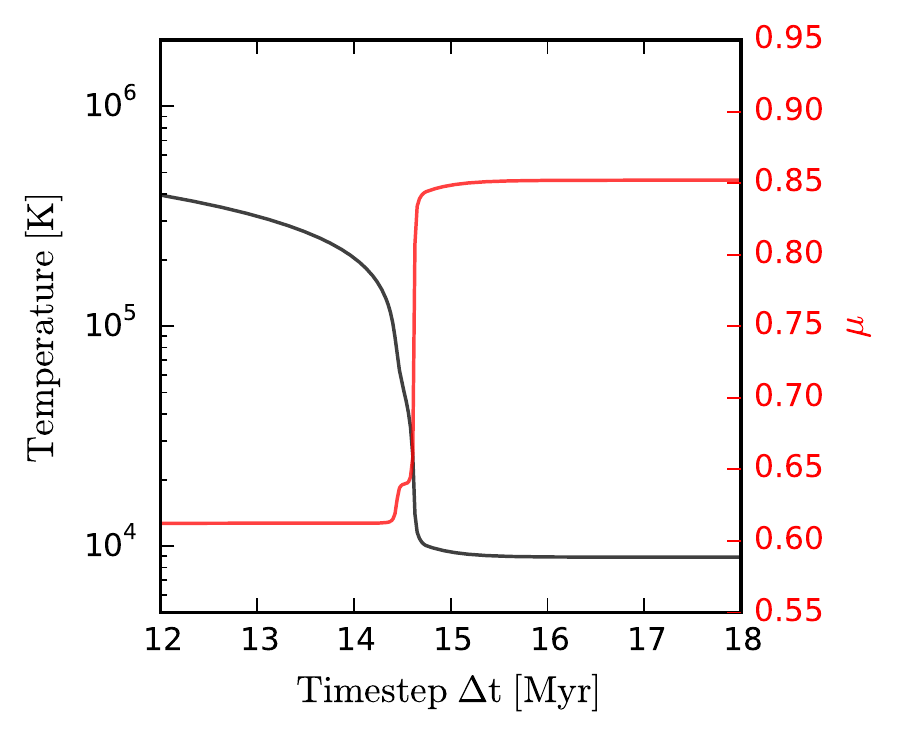}\\
\end{tabular}
\caption{\label{fig:cooling_cell} 
A constant density heating / cooling test of the exact integration scheme. 
The thick black line shows the predicted gas temperature while the thick red 
line shows the mean molecular weight $\mu$. Cooling rate is calculated for 
$0.1Z_{\odot}$ gas and metal-free gas. 
\textit{Top panel}: Predicted gas temperature and $\mu$ from $10$ K as a function of a 
single timestep size $\Delta t$. 
\textit{Middle panel}: Same as the top panel but starting from a temperature of $10^6$ K. 
\textit{Bottom panel}: A zoomed-in view of the temperature and $\mu$ evolution in the middle panel 
between 12 and 18 Myr.}
\end{center}
\end{figure}

\subsection{Validation of the exact integration scheme for gas cooling}

\begin{figure*}
\begin{center}
\begin{tabular}{ccc}
\includegraphics[width=0.33\linewidth]{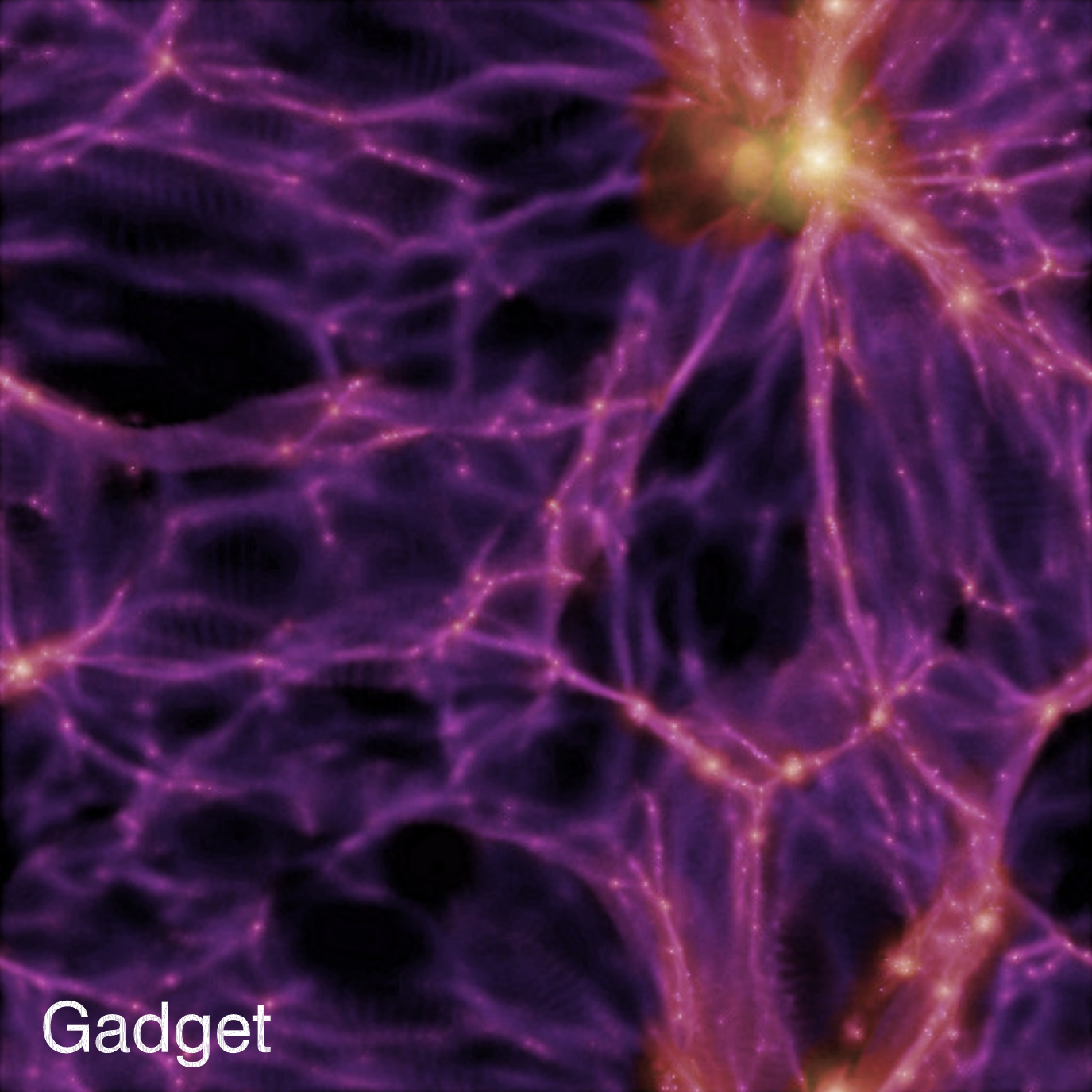}
\includegraphics[width=0.33\linewidth]{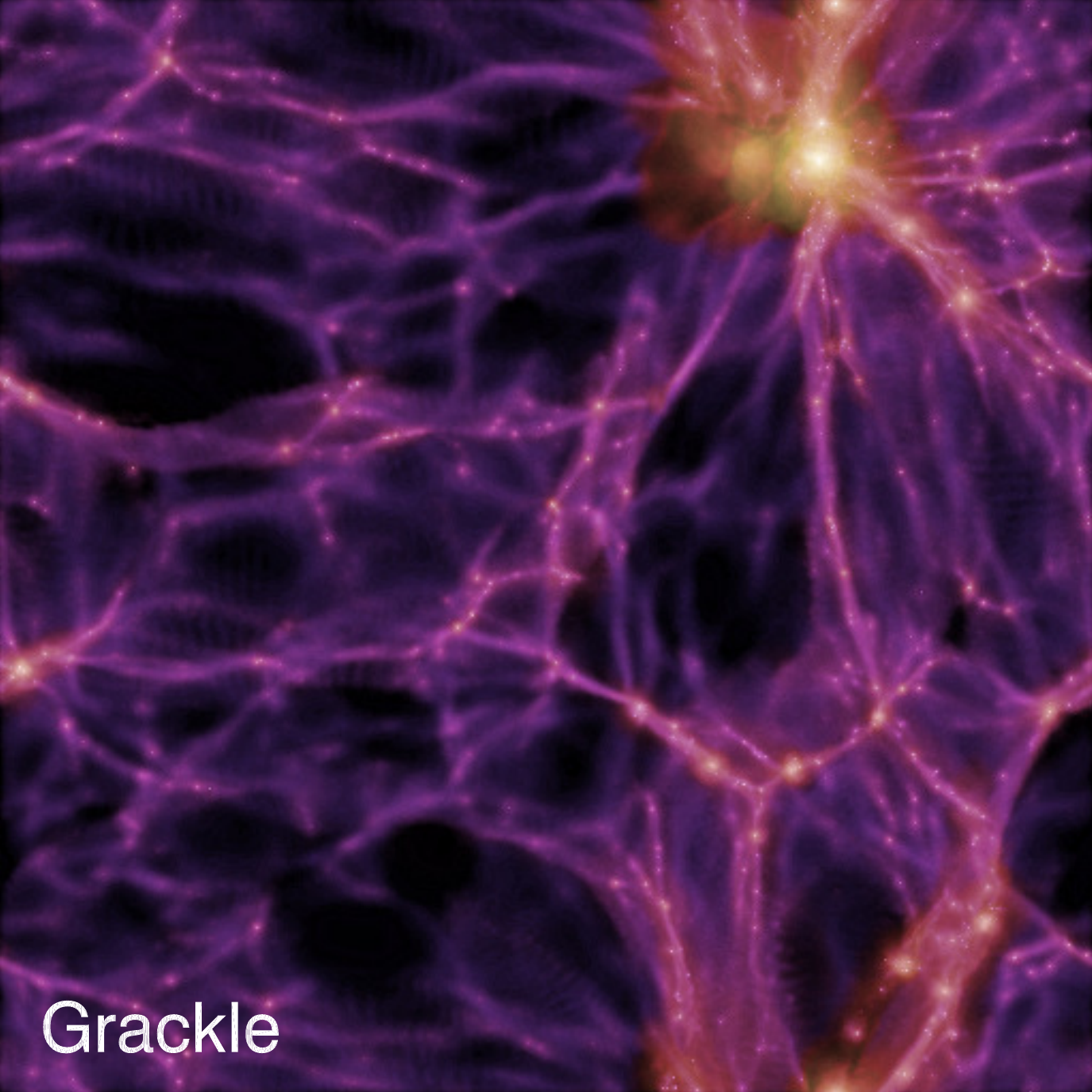}
\includegraphics[width=0.33\linewidth]{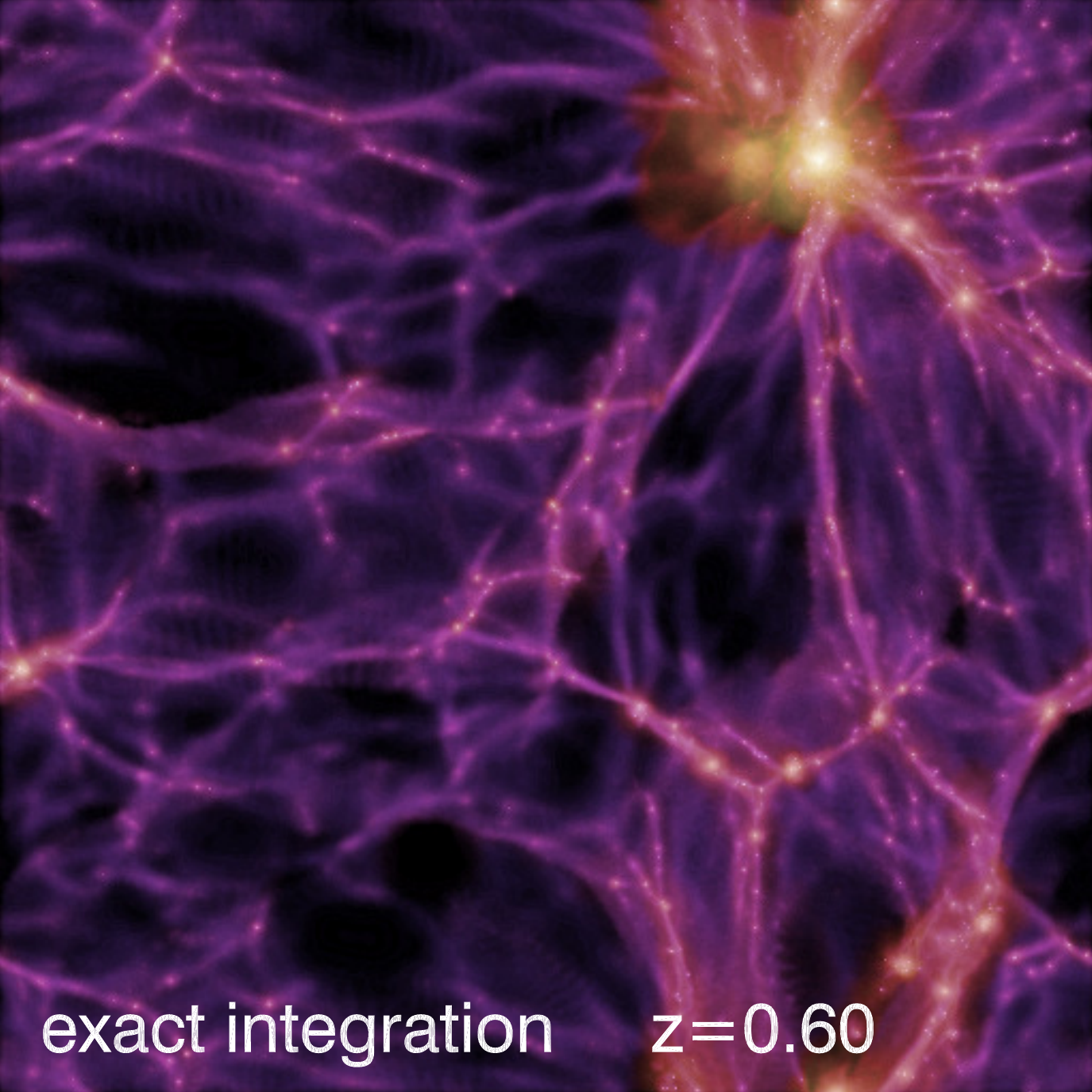}\\
\includegraphics[width=0.33\linewidth]{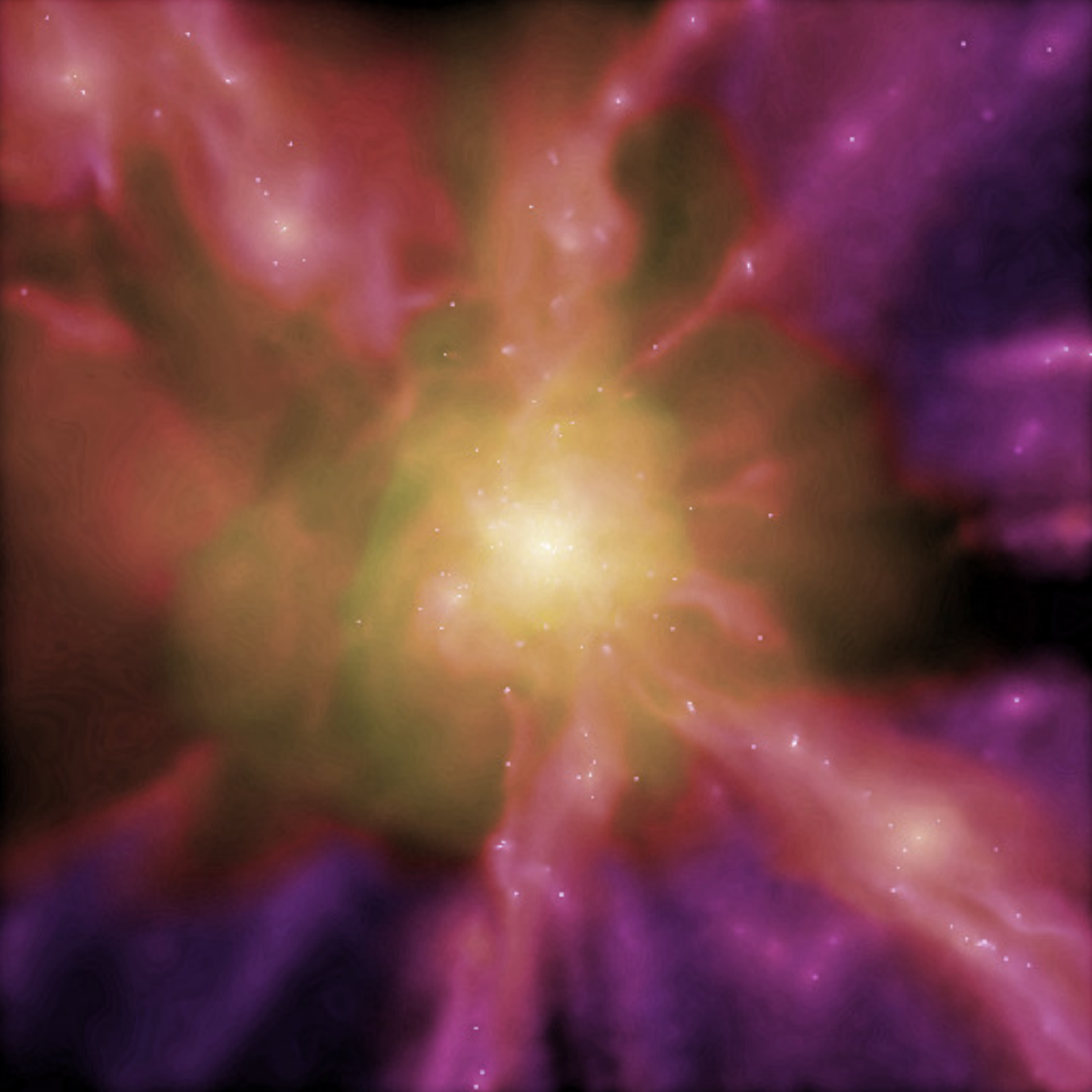}
\includegraphics[width=0.33\linewidth]{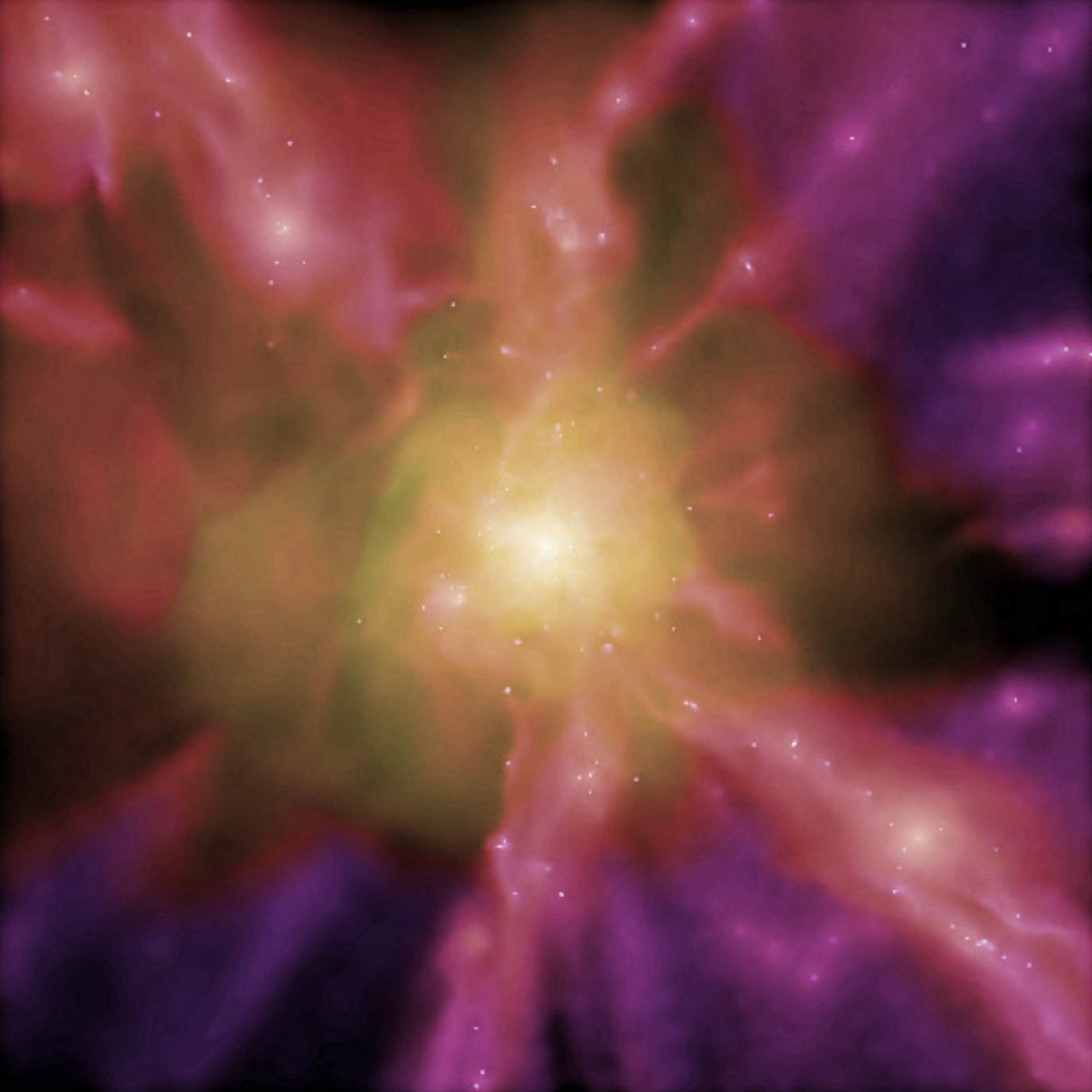}
\includegraphics[width=0.33\linewidth]{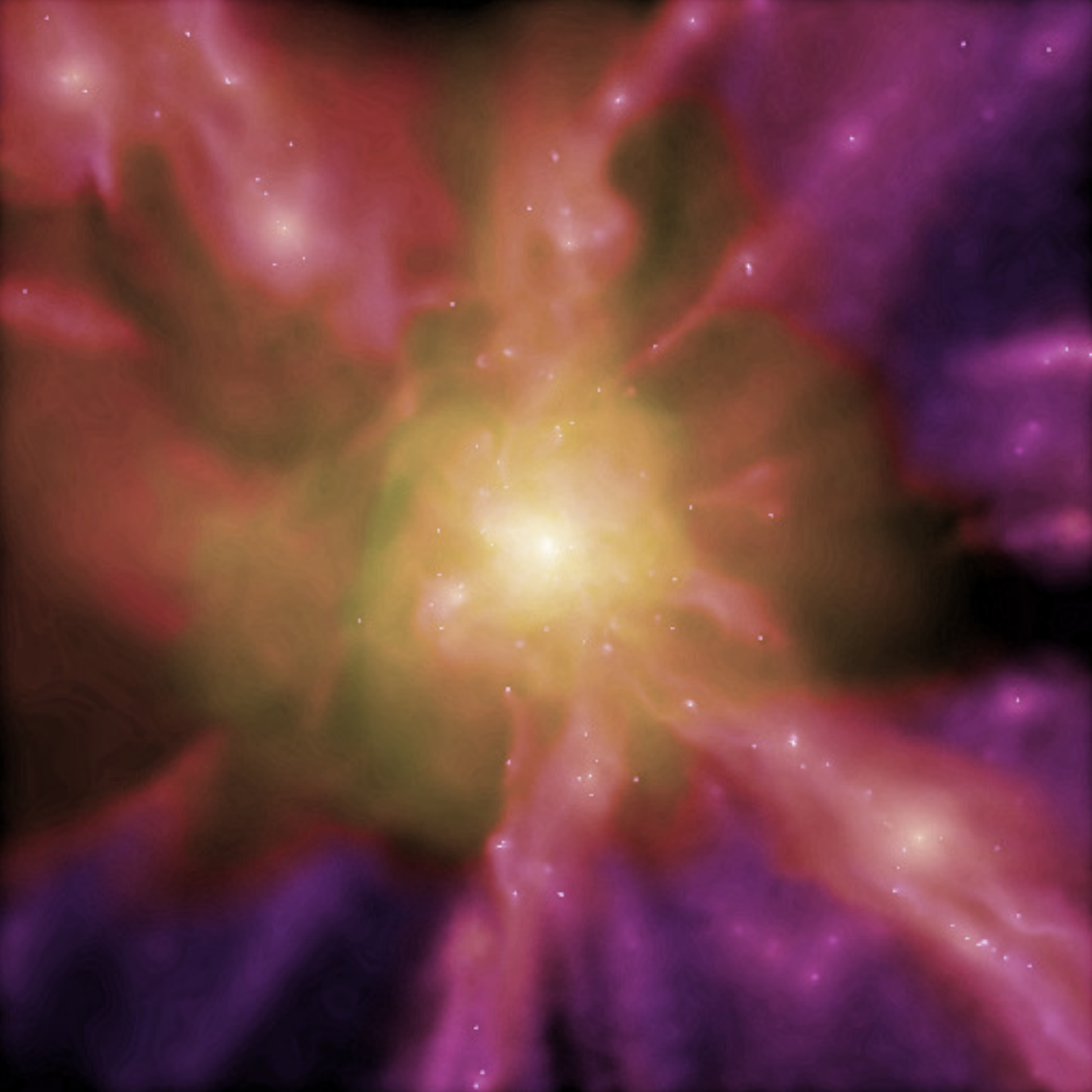}\\
\end{tabular}
\caption{\label{fig:gas_image} 
Gas density distribution in a 30 Mpc$/h$ box cosmological simulation at 
redshift $z = 0.6$ simulated with implicit integration for cooling as used in {\sc Gadget} ({\it left}), 
explicit integration in {\sc Grackle} ({\it middle}) and the exact integration scheme  ({\it right}). The gas temperature is 
color-coded such that cold gas appears in blue while hot gas is yellow. The lower panels
show the zoomed-in view of the gas distribution in the most massive halo. }
\end{center}
\end{figure*}

\begin{figure*}
\begin{center}
\begin{tabular}{cc}
\includegraphics[width=0.33\linewidth]{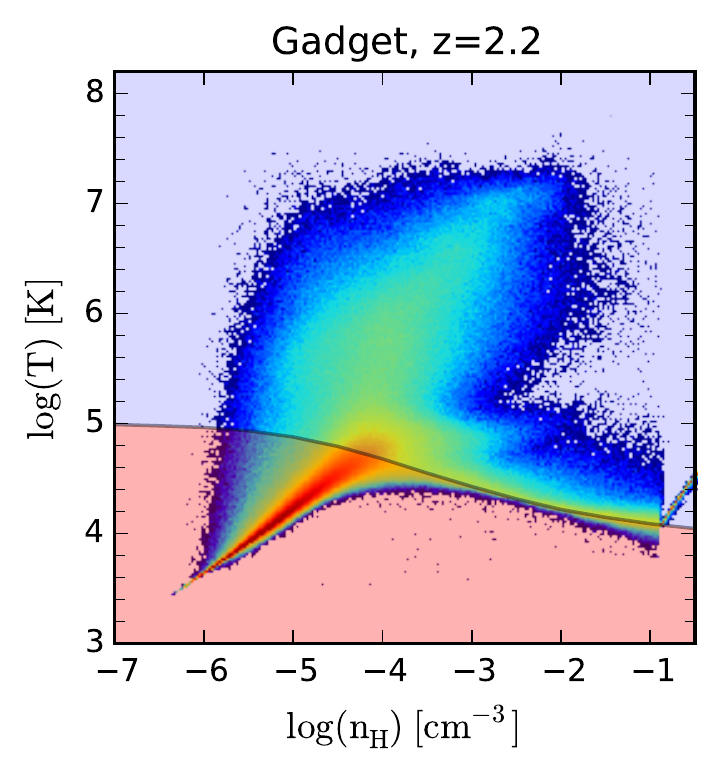}
\includegraphics[width=0.33\linewidth]{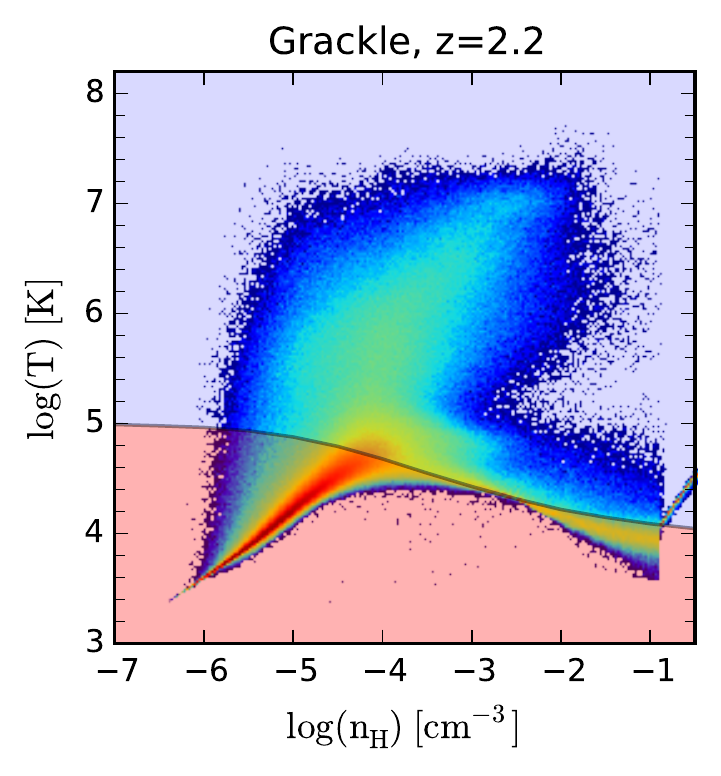}
\includegraphics[width=0.33\linewidth]{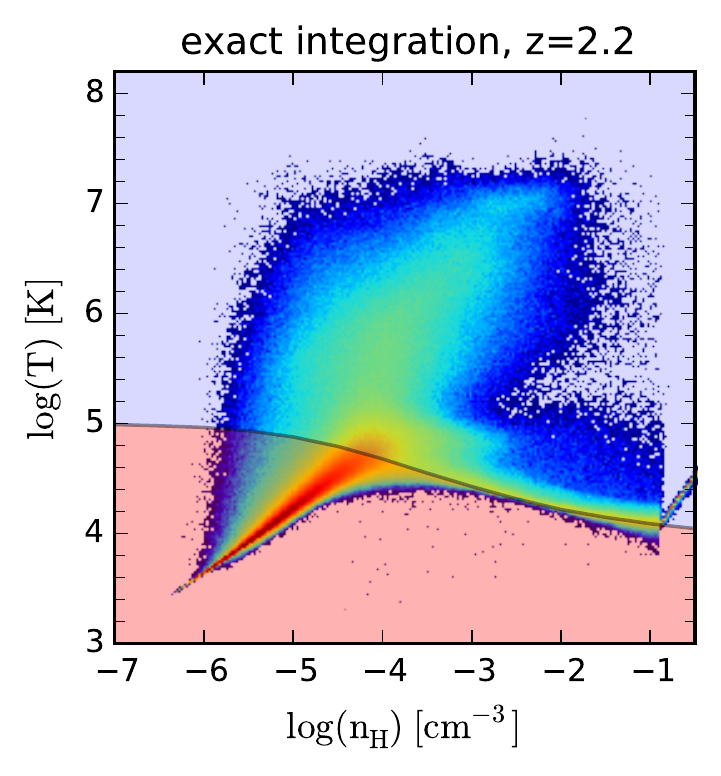}\\
\includegraphics[width=0.33\linewidth]{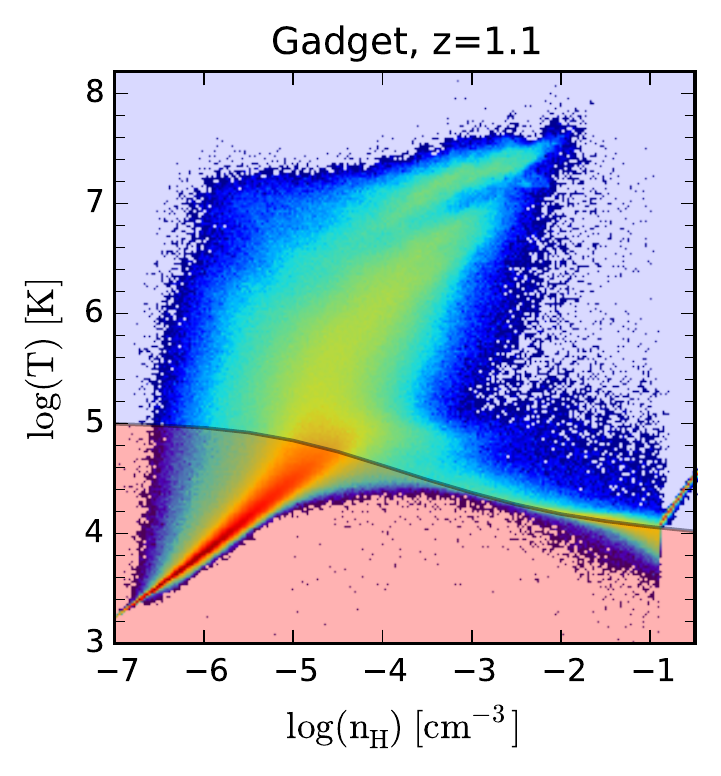}
\includegraphics[width=0.33\linewidth]{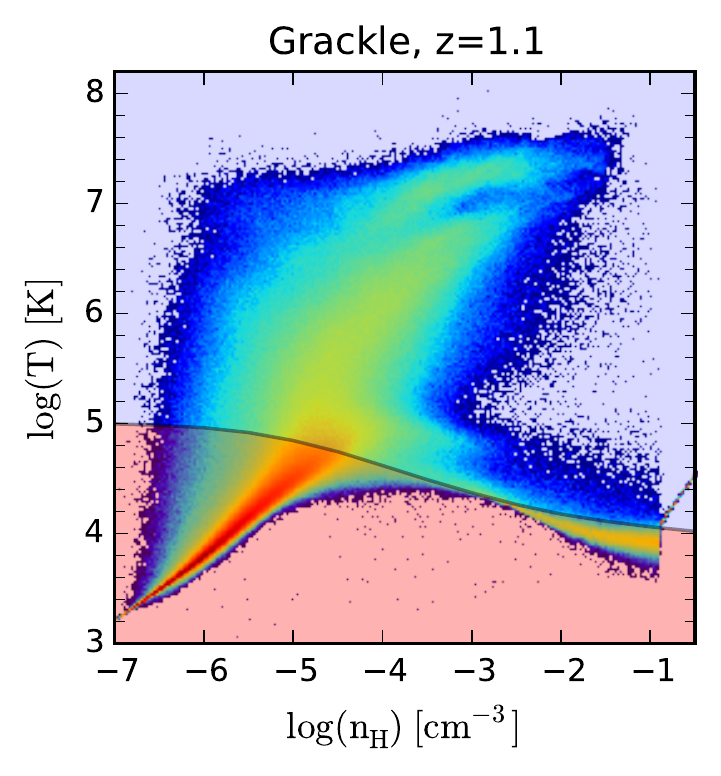}
\includegraphics[width=0.33\linewidth]{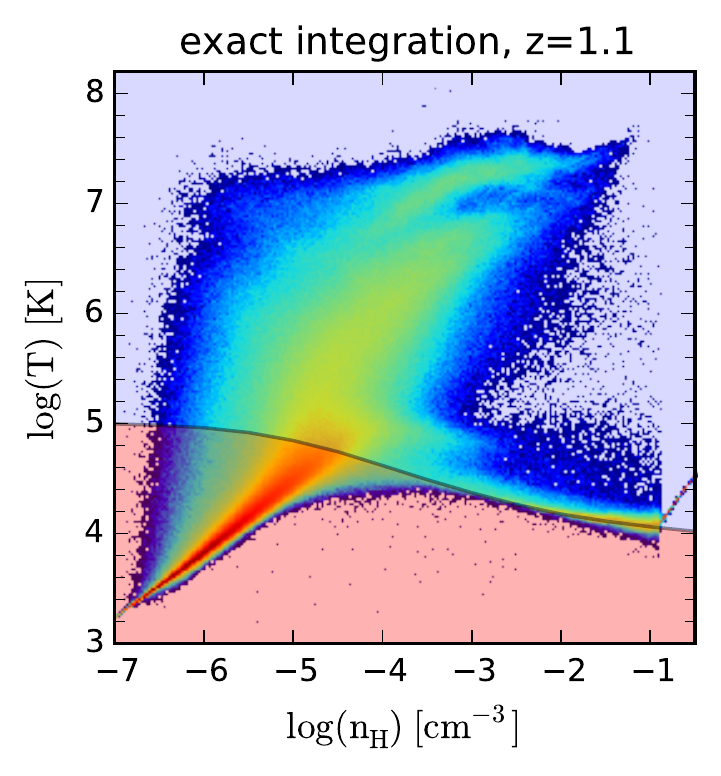}\\
\includegraphics[width=0.33\linewidth]{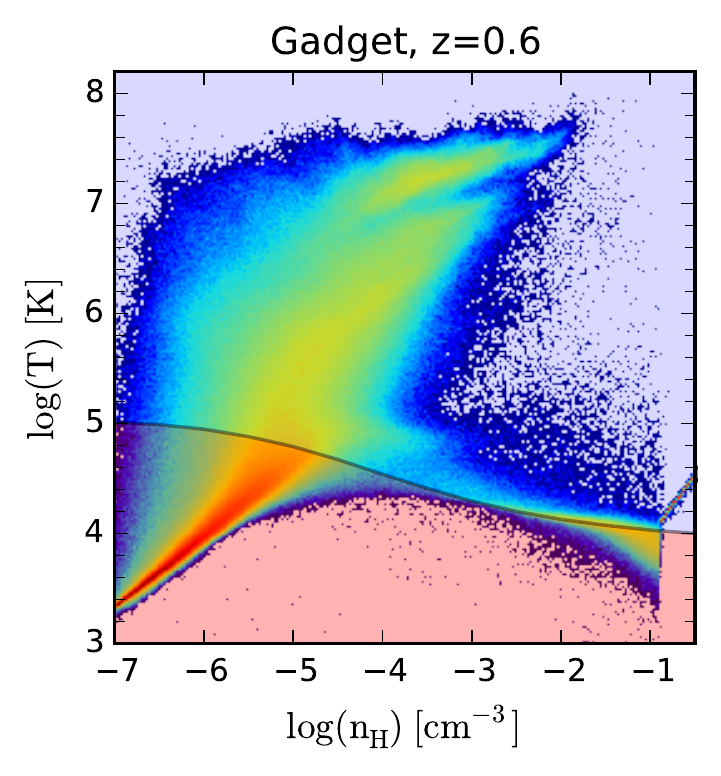}
\includegraphics[width=0.33\linewidth]{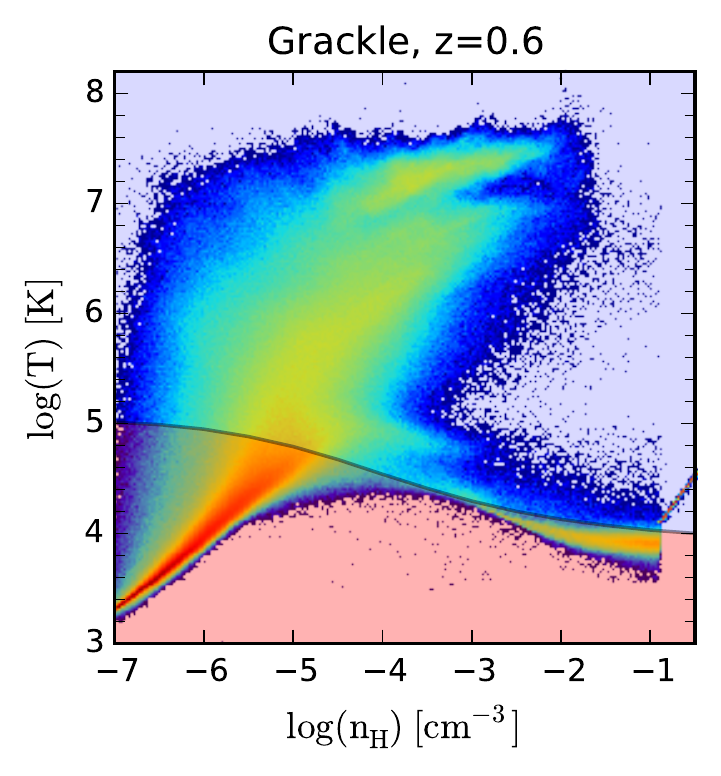}
\includegraphics[width=0.33\linewidth]{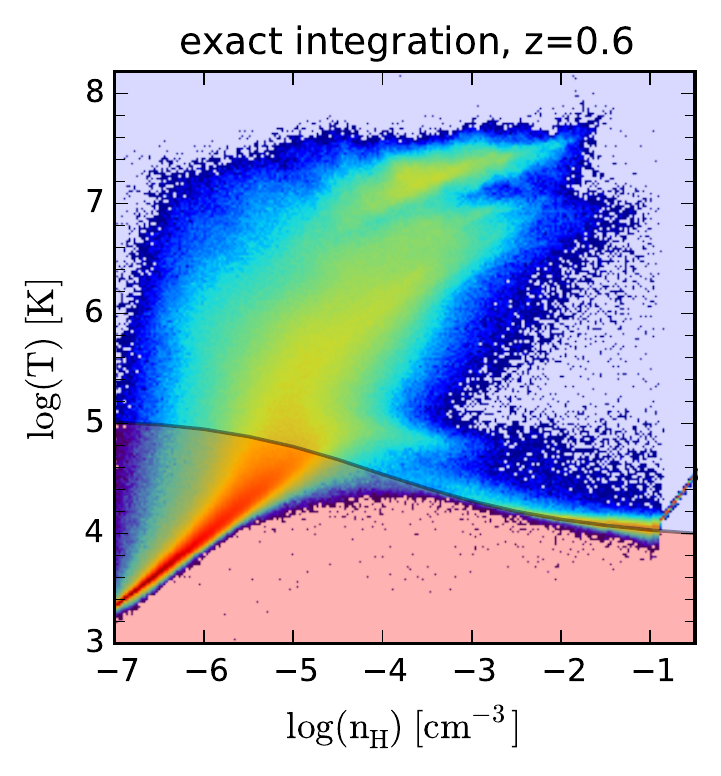}\\
\end{tabular}
\caption{\label{fig:gas_phase} 
Gas density--temperature phase diagram in a 30 Mpc$/h$ box simulated with three integration
schemes. Gas phase diagrams obtained with the three integration methods agree with each other 
in most of the parameter space. However, the temperature for gas with densities in the
$10^{-2} \le n_{\rm H} \le 10^{-1}\, \rm{cm^{-3}}$ range is slightly higher in the exact time 
integration than with the other two approaches, which results in many gas particles with
temperature below the equilibrium line defined as where the cooling and heating rates
are equal. 
The region where gas is subjected to net heating from the UV background is shown in the red band
while the region with net cooling is shown in the blue band.}
\end{center}
\end{figure*}

Before we use this method in actual simulations, we first apply it to
the constant density cooling test similar to \cite{Smith2016}, which
examines the accuracy of the solver over a period of time. A gas cell
is at a number density of $0.1\, \rm{cm}^{-3}$ and an initial
temperature. The gas cooling rate is calculated for $0.1Z_{\odot}$ gas
exposed to the UV background of \cite{Haardt2012} at redshift $z = 0$.

In the top panel of Figure~\ref{fig:cooling_cell}, we show the
gas temperature and the mean molecular weight $\mu$ as a
function of a single timestep $\Delta t$ for a starting
temperature of $10\, \rm{K}$. The gas temperature is expected to rise
up to $\sim$$10^4\, \rm{K}$ due to the heating effect of the UV
background, which completes at $\sim$1 Myr. The predicted
temperature of the metal-free gas case is very similar to
$0.1Z_{\odot}$ gas.

The middle panel shows the gas temperature and $\mu$ for a starting temperature of
$10^6$ K, which is the same as that in \cite{Smith2016}. The gas temperature and 
$\mu$ for $0.1Z_{\odot}$ gas agree perfectly with Figure 3 in \cite{Smith2016}. In the case of 
net cooling, metal-free gas cools more slowly than 
$0.1Z_{\odot}$, which is also subject to metal cooling. 
Note that cooling from $10^5$ to $10^4$ K is a very abrupt phase, 
much shorter than what it takes from $10^3$ to $10^4$ K due to heating.

The above tests show that the exact integration algorithm and
our implementation are both correct in the regimes of net heating and
net cooling. Moreover, the temperature is insensitive to the timestep
size, which contrasts with the behavior of the explicit and implicit
integrations.  The predicted gas temperature shows a sharp drop for
$\Delta t \approx 15\, \rm{Myr} $, which is very difficult to capture
accurately with either explicit or implicit methods if the timestep is
comparable to the cooling time.

The lower panel of Figure~\ref{fig:cooling_cell} shows a zoomed-in
view of the gas temperature and $\mu$ in the cooling test in the
middle panel around the time of the sharp temperature drop.  The
predicted values of temperature and mean molecular weight $\mu$
accurately reflect the features in the cooling curves between
$10^4$ and $10^5$ K (a local minimum between two peaks) within a short
duration of $\sim$0.2 Myr.

\subsection{Cosmological hydrodynamic simulations with different time integration schemes for gas cooling}

\begin{table}
\caption{A comparison of the three time integration schemes for gas used in our cosmological hydrodynamic 
simulations.}
\begin{center}
\begin{tabular}{c|c|c|c}
\toprule
scheme   name         & {\sc Gadget}  & {\sc Grackle} & exact integration      \\\hline
equation                    & Eq.~(\ref{eq:implicit})    & Eq.~(\ref{eq:explicit})  & Eq.~(\ref{eq:deltay3}) \\\hline
order of accuracy      &   first order     &  first order     &       exact                   \\\hline    
sub-cycling                &    no                & yes               & no                             \\       
\bottomrule
\end{tabular}
\end{center}
\label{table:integration_comparison}
\end{table}

Having established that we have a correct implementation of the exact integration, we now
use cosmological hydrodynamic simulations to evaluate the differences between the exact 
integration of gas cooling with an implicit method as in the {\sc Gadget} code and the method 
in {\sc Grackle}. We generate initial conditions in $\Lambda$CDM using the {\sc music} code
\citep{Hahn2011}. A uniform box of $30$ comoving Mpc on each side contains a total number of
$256^3$ gas particles and $256^3$ dark matter particles. Hydrodynamics is evolved using the
moving-finite-mass (MFM) method in the {\sc Gizmo} code. Physical processes include radiative 
cooling and star formation. In order to isolate the effect of radiative cooling due to time integration, 
we do not include any ``kinetic feedback". Only the thermal pressure in the ISM 
\citep{Springel2003} is applied. 

The first simulation uses the exact-integration from the previous section with the cooling curves 
from ``CloudyData\_UVB=FG2011.h5" in {\sc Grackle}. The cooling curves are based on 
{\sc CLOUDY} calculations with the UV background of \cite{Faucher2010}. Our second simulation
adopts a first order backward Euler method implemented in {\sc Gadget}. To solve 
Eq.~(\ref{eq:implicit}), {\sc Gadget} uses a bisection root finding method. The last simulation is 
carried out with {\sc Grackle} in its tabulated cooling mode also with identical cooling curves 
as in the other two simulations. Similar to \cite{Springel2001}, we
apply a restriction such that a particle is only allowed to lose at most 50\% of its internal energy 
in any timestep for all three integration schemes. The numerical parameters controlling the 
gravity calculation, hydrodynamics, and star formation are identical in the three simulations so that any 
differences are caused by the time integration schemes. The differences between the 
methods are listed in Table~\ref{table:integration_comparison}.

In the top panels of Figure~\ref{fig:gas_image}, we show the gas density distribution in the 
simulation box at redshift $z = 0.6$  simulated with the cooling method in {\sc Gadget}, 
{\sc Grackle}, and exact integration scheme in the entire simulation box. Gas temperature 
is color-coded such that cold gas appears in blue while hot gas is in yellow. All 
three simulations produce almost identical structures in the distribution of gas density and
temperature. The bottom panels further show the zoomed-in view of the gas distribution in 
the most massive dark matter halo. Again, differences in the distribution of gas density 
and temperature in the three simulations are not apparent.

\subsubsection{Gas density--temperature phase diagram}
It is not surprising to find that the impact of time integration for gas cooling appears
to be slight due to the fact that gas cooling times in low-density regions and in 
shocked heated regions (galaxy groups) are both very long. 
It is imperative to consider a close inspection of the gas temperature 
in the regime where cooling is efficient. 

To do this, it is a common practice in the field to examine the gas density-temperature diagram. 
To calculate gas temperature from the specific internal energy, we use Eq.~(\ref{eq:temp_u_conversion}) 
with the mean molecular weight in the {\sc Grackle} cooling table as a function of density, 
the specific internal energy, and redshift. This step ensures that the conversion from gas internal energy to 
temperature is accurate\footnote{Since the cooling term is treated with an operator splitting
approach, the update of the mean molecular weight occurs between the two kick operations. 
There is a slight mismatch between the gas internal energy and the mean molecular weight 
in the snapshots.} while the differences in the gas density--temperature phase diagrams are 
solely determined by the time integration. 

In Figure~\ref{fig:gas_phase}, we show the gas density--temperature phase diagrams. In these 
plots, we also emphasize the region where heating due to the UV background dominates 
in red and the region of net cooling is in blue. The boundary between net cooling and 
net heating is shown in a solid gray line, which is an equilibrium where the heating rate 
equals the cooling rate. Since the intensity and the spectrum of the UV background evolve
as a function of redshift, the equilibrium line also varies accordingly. Between $z = 2$ and 
$z = 0.6$, the equilibrium between heating and cooling drops from $\sim$$10^5\, \rm{K}$ for gas
at density
$n_{\rm H} = 10^{-7}\, \rm{cm^{-3}}$ to $\sim$$10^4\, \rm{K}$ for $n_{\rm H} = 10^{-1}\, \rm{cm^{-3}}$. 

Overall, the gas phase diagrams obtained with the three integration schemes agree well
in most of the parameter space. 
However, the temperature for gas in the density range
$10^{-2} \le n_{\rm H} \le 10^{-1}\, \rm{cm^{-3}}$ is slightly higher in the exact time integration 
than the implicit integration and {\sc Grackle}, which contains many gas particles with a temperature 
below $10^4$ K. Physically, this part of the phase diagram corresponds to gas falling into 
dark matter haloes along large scale filaments. 

The gas temperature obtained with the cooling method in {\sc Gadget} shows clear departures from 
equilibrium at $z = 1.1$ and $ z = 0.6$  in $10^{-2}\le n_{\rm H}\le 10^{-1}\, \rm{cm^{-3}}$. 
In particular, the distribution of the gas temperature is skewed below $10^4$ K for gas at densities
around $10^{-1}\, \rm{cm^{-3}}$. Interestingly, the gas temperature follows the equilibrium relation very 
well at $z = 2.2$. For the simulation using {\sc Grackle} cooling, the gas temperature is consistently 
lower than the equilibrium line in the density range $10^{-2} \le n_{\rm H} \le 10^{-1}\, \rm{cm^{-3}}$. 

For comparison, the gas temperature obtained with the exact integration closely follows the 
equilibrium relation at all three redshifts, as shown in Figure~\ref{fig:gas_phase}. 
Moreover, the distribution of gas particles with temperature above and below the equilibrium line
is symmetrical. 

The coupling between the gas dynamics and gas cooling in the hydrodynamic simulations 
adds more complexity in the interpretation of the departure from the equilibrium temperature with 
{\sc Gadget} and {\sc Grackle} in the phase diagram compared to the constant density cooling 
test in Figure~\ref{fig:cooling_cell_grackle}. As the gas in the cosmic web enters into the gravitational 
potential of galaxies, it cools towards the equilibrium temperature but gas density 
continuously increases which moves the equilibrium temperature lower. 
The increasingly higher density could introduce a bias which amplifies 
the numerical error towards a lower gas temperature as the gas density can even reach 
a higher value.

In the case of {\sc Grackle}, another bias towards a lower temperature is also present. 
The cooling rate and the heating rate are not symmetric around the equilibrium temperature.
This bias is present in Figure~\ref{fig:cooling_cell} as it takes much shorter time to
cool the gas from $10^5$ to $\sim10^4$ K than heat it from $10^3$ to $\sim10^4$ K. 
Once the temperature oscillates around the equilibrium, one would expect to find 
more gas particles in net heating than in net cooling.

\subsubsection{$M_{\rm DM}$--$M_{\rm stellar}$ relation}
 
\begin{figure}
\begin{center}
\begin{tabular}{c}
\includegraphics[trim=0.25cm 0.1cm 0.0 0.2cm,clip,width=\linewidth]{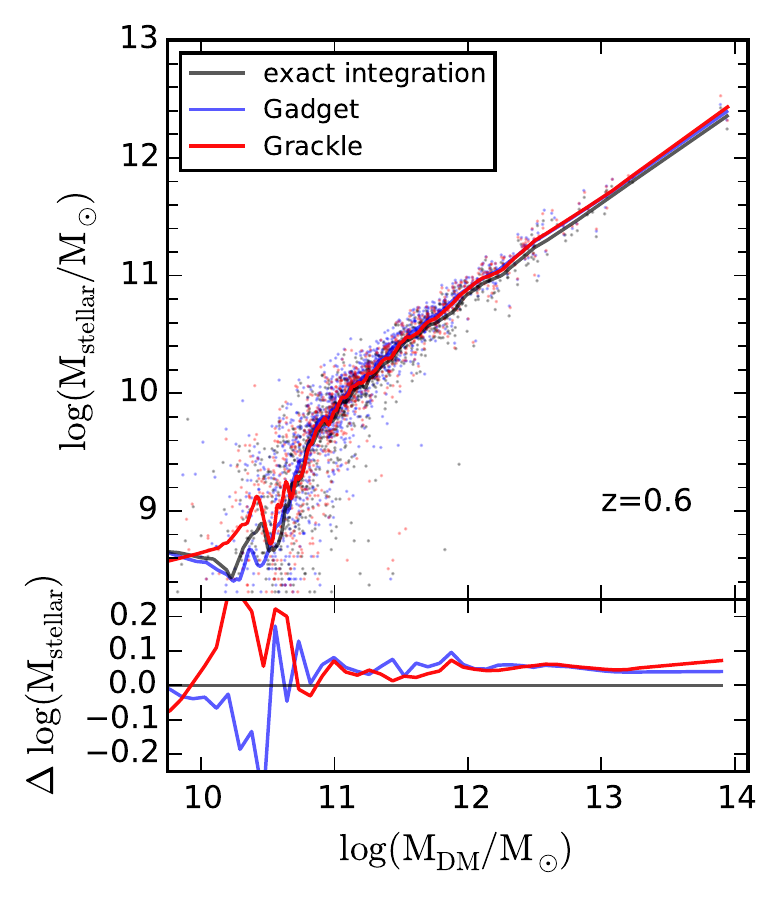}\\
\end{tabular}
\caption{\label{fig:mdm_mstar_comparison} A comparison of stellar mass--dark matter mass 
relation at $z = 0.6$ between the three simulations. The solid lines are obtained using 
a locally weighted regression. The differences in the  stellar masses for a given dark matter 
halo mass in the three simulations is less than $0.1$ dex. The difference in the stellar 
mass in the three simulations is consistent between dark matter haloes more massive 
than $10^{11} \Msun$.}
\end{center}
\end{figure}

While gas cooling and star formation proceed almost the same in the three simulations, 
some quantitative differences are to be expected in the stellar mass using different time integration 
schemes. Overall, we find there are actually minor differences in the three star formation histories 
which are reflected in the total stellar mass produced in the simulation box. The stellar mass obtained 
with {\sc Grackle} cooling is slightly higher than that obtained with the cooling method in 
{\sc Gadget}. The exact integration scheme produces the least stellar mass among the 
three. Overall, the differences in total stellar mass are small,
at the level of tens of percents.
Nevertheless, this is an encouraging and welcoming feature of the exact 
integration scheme as it suffers least from the over-cooling problem.

We look further at the stellar mass in individual dark matter haloes. In 
Figure~\ref{fig:mdm_mstar_comparison} 
we compare the stellar mass--dark matter mass relations at $z = 0.6$ with different time integration
schemes. In this plot, the scattered data points show the dark matter halo mass and the stellar mass for
each FOF halo and the solid lines are locally weighted regression results. 

We use the stellar mass from the exact integration $M_{\rm stellar}^{\rm exact}$ as 
a baseline for comparison. The differences in stellar mass is shown in the lower part of 
Figure~\ref{fig:mdm_mstar_comparison} in terms of
$\Delta \log(M_{\rm stellar}) = \log(M_{\rm stellar})  - \log(M_{\rm stellar}^{\rm exact})$. 
The stellar mass for a given dark matter halo mass in the three simulations 
agrees to within $0.1$ dex.  Nevertheless, the differences in
stellar mass between the exact approach and the other integrators is
systematic
for dark matter haloes more massive than $10^{11} \Msun$, with the exact
technique yielding the lowest masses. 
Since the peak of star formation efficiencies occurs at $\sim$$10^{12} \Msun$ 
halos, the exact integration scheme could at least alleviate the quantitive need for 
supernovae and/or AGN feedback required in the current hydrodynamic simulations 
\citep{Vogelsberger2013, Pillepich2017, Weinberger2017}.

\subsubsection{Is the cooling time well resolved in our simulations?}

\begin{figure}
\begin{center}
\begin{tabular}{c}
\includegraphics[trim=0.25cm 0.2cm 0.0 0.0,clip,width=\linewidth]{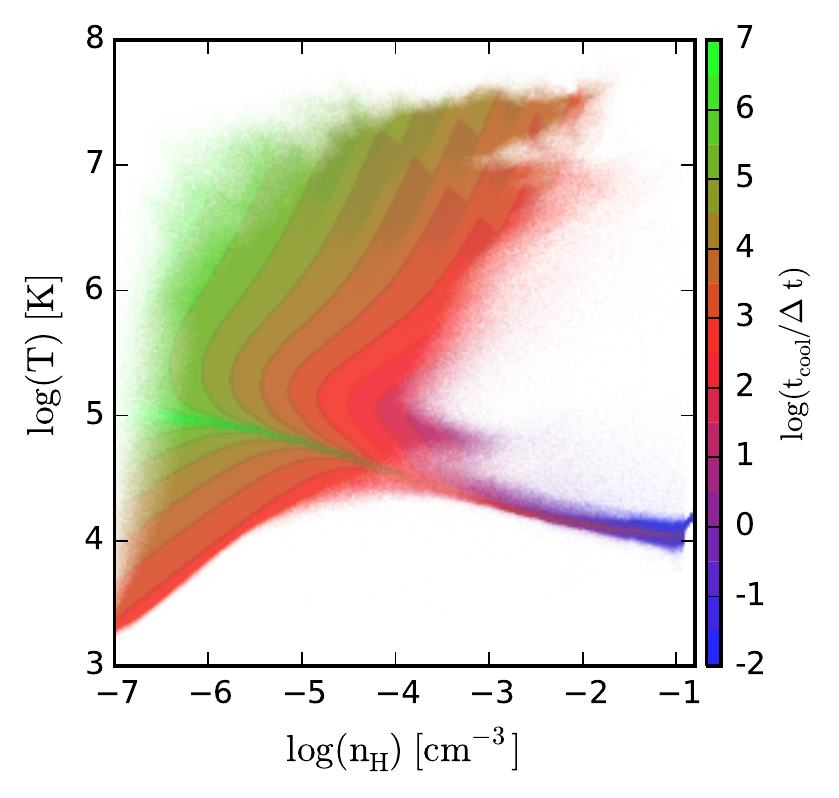}\\
\end{tabular}
\caption{\label{fig:coolingtime_ratio_distribution} The distribution of the ratio between 
the cooling time and the dynamical time for each particle in the simulation with the
exact time integration. In the filaments, a population of gas particles with 
$t_{\rm cool}/{\Delta t} \ll 1$ is also present. See text for more details. }
\end{center}
\end{figure}

Inevitably, our hydrodynamic simulations are constrained by both 
mass and force resolutions. Another resolution effect, not discussed 
often in the literature, is worth some discussion here. Even with a 
set of well-defined cooling curves and accurate hydrodynamic solvers 
available today, the large range of cooling times in a typical cosmological 
simulation itself presents a significant numerical challenge. Short gas 
cooling times on the order of $10^5$ yr in \cite{Wiersma2009} indicates 
that there could be certain gas phases not well-resolved in current 
cosmological simulations.

In numerical simulations, many codes do not apply any restrictions on timesteps due to radiative
cooling. We use the system timestep $\Delta t$ from the simulations, which is mostly determined
by the CFL condition \citep{Saitoh2010}. In our tests, we have applied a restriction that each particle 
could at most lose 50\% its thermal energy in a single step. We would expect this constraint 
to become fully redundant once the shortest cooling time is fully resolved.

Figure~\ref{fig:coolingtime_ratio_distribution} 
shows the distribution of the ratio between the cooling time and the timestep $\Delta t$ for each 
particle at $z = 0.6$ with the exact time integration. We use a simple mapping from the ratios
to colors of gas particles as shown in the color-bar. A very long cooling time with respect to $\Delta t$ 
is shown in green while very short cooling times are shown in blue. The distribution of the ratio
between the two time timescales spans almost ten orders of magnitude. 

For the shock-heated region, the low density gas with temperature above $10^6$ K, 
the cooling time is much longer than the timestep. Gas particles near the equilibrium where 
the cooling rate equals the heating rate also show long cooling times, which is expected since 
the net cooling rate approaches zero. This is why the phase-space diagrams in the three 
simulations appear similar to each other. 

The equilibrium where the cooling and heating rates are equal shows up in 
Figure~\ref{fig:coolingtime_ratio_distribution} as a prominent green line
at $\log(T)\sim 5$  and $n_{\rm H}\sim10^{-7}\, \rm{cm^{-3}}$. For gas particles in the filaments
($\log(T)\sim 4$ and $10^{-2} \le n_{\rm H} \le 10^{-1}\, \rm{cm^{-3}}$), the gas temperature lies 
around the equilibrium in the exact integration scheme. 
On the other hand, $t_{\rm cool}/\Delta t$ in this phase does not exceed 
$10^5$ as the green line terminates around $n_{\rm H} = 10^{-4}\, \rm{cm^{-3}}$.
This is caused by more dynamical 
states in the filaments where it is difficult for gas particles to lie perfectly on the equilibrium 
as they fall into the gravitational potential and interact with ambient gas. 
Above and below the equilibrium line, there is a population of gas particles with 
$t_{\rm cool}/\Delta t \sim 0.01$. This short timescale for gas cooling translates into 
a very large single timestep, which is challenging for both {\sc Gadget} and {\sc Grackle}
cooling methods to maintain their accuracy. This is also why there are differences in 
the gas density--temperature phase diagrams in the filaments. 
Note that temperature around the equilibrium line with the exact
cooling scheme could still contain substantial numerical errors due to the fact
that none of the three integration schemes resolve the cooling time well. Ultra-high
resolution simulations \cite[e.g.,][]{Nelson2016} are essentially needed to better 
understand the gas thermal and dynamical states in this phase.

\section{Discussion}
\label{sec:discussion} 

The exact time integration scheme we have studied in this work can be easily extended to more
sophisticated applications to include self-shielding \citep{Rahmati2013}, or
a local radiation field \citep{Vogelsberger2013, Gnedin2012} or molecular hydrogen 
($\rm{H_2}$) cooling, which requires one to generate cooling tables using photo-ionization 
codes such as {\sc CLOUDY}. On the other hand, this method is difficult to apply to 
non-equilibrium cooling. 

\begin{table}
\caption{A comparison of wall-time spent by three time integration schemes.}
\begin{center}
\begin{tabular}{c|c|c|c}
\toprule
scheme name           & {\sc Gadget}  & {\sc Grackle} & exact   integration      \\\hline
average wall-time \tablefootnote{The average time is taken directly from the
measurement in raw wall-time statistics. Due to the differences between the 
actual nodes carrying out the simulations, a more accurate comparison would 
be the fraction of each procedure within a single timestep.} & 178 s  & 75 s &  87 s     \\\hline
gravity                            & 20.8 s           &   18.9 s           &  20.1 s         \\
(tree+PM)                      & 12\%            &   25\%            &  23\%    \\\hline
                                      & 34.2 s           &   29.7 s          &  33.9 s         \\
hydrodynamics              &19\%     &   40\%            &  39\%        \\\hline
domain                          & 8.1 s           &   9.3 s            &  9.6 s      \\
decomposition              & 5\%              &12\%       &  11\%      \\\hline
cooling                          & {\bf 107.6  s}  & {\bf 7.1 s}  &  {\bf 9.4 s}        \\
and star formation        & {\bf 60\%}  & {\bf 9\%}  &  {\bf 11\%} \\
\bottomrule
\end{tabular}
\end{center}
\label{table:timings}
\end{table}

The exact time integration is efficient when compared to the implicit integration which invokes 
multiple calls to interpolate  gas cooling rates. In Table~\ref{table:timings}, we compare the 
wall-time statistics of the three integration schemes used in  our simulations. 
The wall-time spent by gravity, hydrodynamics, and domain decomposition in the three 
simulations are almost the same. The major difference is in gas cooling: the exact 
integration consumes one tenth as much as the implicit scheme while the latter is significantly slowed
by multiple table look-ups and interpolations, which we have not optimized, 
during its root finding step. In theory, the cost of the exact integration scales as 
$N_{\rm gas} N_{\rm grid}$, the product of the number of gas particles $N_{\rm gas}$ 
and the number of temperature grid points $N_{\rm grid}$. 

This time integration scheme might also help to resolve gas temperature for studies of cooling 
emission \citep[][]{Haiman2000, Fardal2001ApJ, Faucher2010a}. Because gas 
at temperatures around $\sim$$10^4$ K in the
current simulations is not well-resolved due to the short cooling time and large intrinsic 
variations in the cooling timescale, as we showed in the previous section, large uncertainties in the
cooling emission could be present as the Lyman-$\alpha$ emissivity changes exponentially around  
$\sim$$10^4$ K \citep{Faucher2010}. 

While the differences in the stellar mass produced by the three time integration schemes for gas
cooling are relatively small ($<0.1$ dex), the galaxy formation model we used in our simulation does 
not resolve the ISM dynamics once the gas is above the threshold density  
(just slightly above $0.1\, \rm{cm^{-3}}$) for star formation. As we see from the cooling timescale 
distribution in Figure~\ref{fig:coolingtime_ratio_distribution}, gas cooling for most of the particles 
in our simulations is well-resolved except for gas more dense than $n_{\rm H} \sim 0.01\ {\rm cm^{-3}}$. 
The limited range in gas densities in our simulations where 
there are large errors in gas cooling could be the reason 
why only relatively 
small differences in the stellar mass are found. One might expect that larger differences due 
to different time integration schemes could be present, as shown in the right panel of
Figure~\ref{fig:cooling_cell_grackle},  for simulations aiming at explicitly resolving 
the ISM physics. 

\section{Conclusions}
\label{sec:conclusions} 

In this paper, we have extended the method of exact time integration by \cite{Townsend2009} to use 
a redshift-dependent cooling table with a UV background. We have applied this method to a
cosmological hydrodynamic simulation and compared its performance with other time integration
schemes. The impact of numerical errors with the current integration in gas cooling on galaxy 
formation appears slight but is present nevertheless. Based on our findings, we
argue that the exact time integration for gas cooling is attractive because

\begin{enumerate}
\item[(1)] It is able to map the specific internal energy to the timestep while it is also
insensitive to the timestep. 
Errors are only from interpolation of the cooling table.
\item[(2)] It is easy to implement and efficient to execute. 
\item[(3)] It is also very flexible in that it can be adapted to
different cooling tables which incorporate different UV
backgrounds, metal-dependent cooling, and other physical processes such as including local
radiation fields.  
\item[(4)] It could in principle moderate the quantitive need for supernovae and/or AGN feedback
in the current models.
\end{enumerate}

\section*{Acknowledgements}
We thank the referee for his/her thorough review 
with very helpful comments on earlier drafts of the manuscript.
This work is supported by NSF grants AST-0965694, AST-1009867,
and AST-1412719. We acknowledge the Institute For CyberScience at The Pennsylvania 
State University for providing computational resources and services that have contributed 
to the research results reported in this paper. The Institute for Gravitation and the Cosmos
is supported by the Eberly College of Science and the Office of the Senior Vice President for 
Research at the Pennsylvania State University. Some of the computations in this paper were 
run on the Odyssey cluster supported by the FAS Division of Science, Research Computing 
Group at Harvard University.


\begin{thebibliography}{}
\makeatletter
\relax
\def\mn@urlcharsother{\let\do\@makeother \do\$\do\&\do\#\do\^\do\_\do\%\do\~}
\def\mn@doi{\begingroup\mn@urlcharsother \@ifnextchar [ {\mn@doi@}
  {\mn@doi@[]}}
\def\mn@doi@[#1]#2{\def\@tempa{#1}\ifx\@tempa\@empty \href
  {http://dx.doi.org/#2} {doi:#2}\else \href {http://dx.doi.org/#2} {#1}\fi
  \endgroup}
\def\mn@eprint#1#2{\mn@eprint@#1:#2::\@nil}
\def\mn@eprint@arXiv#1{\href {http://arxiv.org/abs/#1} {{\tt arXiv:#1}}}
\def\mn@eprint@dblp#1{\href {http://dblp.uni-trier.de/rec/bibtex/#1.xml}
  {dblp:#1}}
\def\mn@eprint@#1:#2:#3:#4\@nil{\def\@tempa {#1}\def\@tempb {#2}\def\@tempc
  {#3}\ifx \@tempc \@empty \let \@tempc \@tempb \let \@tempb \@tempa \fi \ifx
  \@tempb \@empty \def\@tempb {arXiv}\fi \@ifundefined
  {mn@eprint@\@tempb}{\@tempb:\@tempc}{\expandafter \expandafter \csname
  mn@eprint@\@tempb\endcsname \expandafter{\@tempc}}}

\bibitem[\protect\citeauthoryear{{Bauer} \& {Springel}}{{Bauer} \&
  {Springel}}{2012}]{Bauer2012}
{Bauer} A.,  {Springel} V.,  2012, \mn@doi [\mnras]
  {10.1111/j.1365-2966.2012.21058.x}, \href
  {http://adsabs.harvard.edu/abs/2012MNRAS.423.2558B} {423, 2558}

\bibitem[\protect\citeauthoryear{{Birnboim} \& {Dekel}}{{Birnboim} \&
  {Dekel}}{2003}]{Birnboim2003}
{Birnboim} Y.,  {Dekel} A.,  2003, \mn@doi [\mnras]
  {10.1046/j.1365-8711.2003.06955.x}, \href
  {http://adsabs.harvard.edu/abs/2003MNRAS.345..349B} {345, 349}

\bibitem[\protect\citeauthoryear{{Blumenthal}, {Faber}, {Primack}  \&
  {Rees}}{{Blumenthal} et~al.}{1984}]{Blumenthal1984}
{Blumenthal} G.~R.,  {Faber} S.~M.,  {Primack} J.~R.,   {Rees} M.~J.,  1984,
  \mn@doi [\nat] {10.1038/311517a0}, \href
  {http://adsabs.harvard.edu/abs/1984Natur.311..517B} {311, 517}

\bibitem[\protect\citeauthoryear{{Fardal}, {Katz}, {Gardner}, {Hernquist},
  {Weinberg}  \& {Dav{\'e}}}{{Fardal} et~al.}{2001}]{Fardal2001ApJ}
{Fardal} M.~A.,  {Katz} N.,  {Gardner} J.~P.,  {Hernquist} L.,  {Weinberg}
  D.~H.,   {Dav{\'e}} R.,  2001, \mn@doi [\apj] {10.1086/323519}, \href
  {http://adsabs.harvard.edu/abs/2001ApJ...562..605F} {562, 605}

\bibitem[\protect\citeauthoryear{{Faucher-Gigu{\`e}re}, {Kere{\v s}},
  {Dijkstra}, {Hernquist}  \& {Zaldarriaga}}{{Faucher-Gigu{\`e}re}
  et~al.}{2010a}]{Faucher2010}
{Faucher-Gigu{\`e}re} C.-A.,  {Kere{\v s}} D.,  {Dijkstra} M.,  {Hernquist} L.,
    {Zaldarriaga} M.,  2010a, \mn@doi [\apj] {10.1088/0004-637X/725/1/633},
  \href {http://adsabs.harvard.edu/abs/2010ApJ...725..633F} {725, 633}

\bibitem[\protect\citeauthoryear{{Faucher-Gigu{\`e}re}, {Kere{\v s}},
  {Dijkstra}, {Hernquist}  \& {Zaldarriaga}}{{Faucher-Gigu{\`e}re}
  et~al.}{2010b}]{Faucher2010a}
{Faucher-Gigu{\`e}re} C.-A.,  {Kere{\v s}} D.,  {Dijkstra} M.,  {Hernquist} L.,
    {Zaldarriaga} M.,  2010b, \mn@doi [\apj] {10.1088/0004-637X/725/1/633},
  \href {http://adsabs.harvard.edu/abs/2010ApJ...725..633F} {725, 633}

\bibitem[\protect\citeauthoryear{{Ferland} et~al.,}{{Ferland}
  et~al.}{2013}]{Cloudy}
{Ferland} G.~J.,  et~al., 2013, \rmxaa, \href
  {http://adsabs.harvard.edu/abs/2013RMxAA..49..137F} {49, 137}

\bibitem[\protect\citeauthoryear{{Gnedin} \& {Hollon}}{{Gnedin} \&
  {Hollon}}{2012}]{Gnedin2012}
{Gnedin} N.~Y.,  {Hollon} N.,  2012, \mn@doi [\apjs]
  {10.1088/0067-0049/202/2/13}, \href
  {http://adsabs.harvard.edu/abs/2012ApJS..202...13G} {202, 13}

\bibitem[\protect\citeauthoryear{{Haardt} \& {Madau}}{{Haardt} \&
  {Madau}}{2012}]{Haardt2012}
{Haardt} F.,  {Madau} P.,  2012, \mn@doi [\apj] {10.1088/0004-637X/746/2/125},
  \href {http://adsabs.harvard.edu/abs/2012ApJ...746..125H} {746, 125}

\bibitem[\protect\citeauthoryear{{Hahn} \& {Abel}}{{Hahn} \&
  {Abel}}{2011}]{Hahn2011}
{Hahn} O.,  {Abel} T.,  2011, \mn@doi [\mnras]
  {10.1111/j.1365-2966.2011.18820.x}, \href
  {http://adsabs.harvard.edu/abs/2011MNRAS.415.2101H} {415, 2101}

\bibitem[\protect\citeauthoryear{{Haiman}, {Spaans}  \& {Quataert}}{{Haiman}
  et~al.}{2000}]{Haiman2000}
{Haiman} Z.,  {Spaans} M.,   {Quataert} E.,  2000, \mn@doi [\apjl]
  {10.1086/312754}, \href {http://adsabs.harvard.edu/abs/2000ApJ...537L...5H}
  {537, L5}

\bibitem[\protect\citeauthoryear{{Hernquist} \& {Katz}}{{Hernquist} \&
  {Katz}}{1989}]{Hernquist1989}
{Hernquist} L.,  {Katz} N.,  1989, \mn@doi [\apjs] {10.1086/191344}, \href
  {http://adsabs.harvard.edu/abs/1989ApJS...70..419H} {70, 419}

\bibitem[\protect\citeauthoryear{{Hopkins}}{{Hopkins}}{2015}]{Gizmo}
{Hopkins} P.~F.,  2015, \mn@doi [\mnras] {10.1093/mnras/stv195}, \href
  {http://adsabs.harvard.edu/abs/2015MNRAS.450...53H} {450, 53}

\bibitem[\protect\citeauthoryear{{Kere{\v s}}, {Katz}, {Weinberg}  \&
  {Dav{\'e}}}{{Kere{\v s}} et~al.}{2005}]{Keres2005}
{Kere{\v s}} D.,  {Katz} N.,  {Weinberg} D.~H.,   {Dav{\'e}} R.,  2005, \mn@doi
  [\mnras] {10.1111/j.1365-2966.2005.09451.x}, \href
  {http://adsabs.harvard.edu/abs/2005MNRAS.363....2K} {363, 2}

\bibitem[\protect\citeauthoryear{{Kere{\v s}}, {Vogelsberger}, {Sijacki},
  {Springel}  \& {Hernquist}}{{Kere{\v s}} et~al.}{2012}]{Keres2012}
{Kere{\v s}} D.,  {Vogelsberger} M.,  {Sijacki} D.,  {Springel} V.,
  {Hernquist} L.,  2012, \mn@doi [\mnras] {10.1111/j.1365-2966.2012.21548.x},
  \href {http://adsabs.harvard.edu/abs/2012MNRAS.425.2027K} {425, 2027}

\bibitem[\protect\citeauthoryear{{Nelson}, {Vogelsberger}, {Genel}, {Sijacki},
  {Kere{\v s}}, {Springel}  \& {Hernquist}}{{Nelson} et~al.}{2013}]{Nelson2013}
{Nelson} D.,  {Vogelsberger} M.,  {Genel} S.,  {Sijacki} D.,  {Kere{\v s}} D.,
  {Springel} V.,   {Hernquist} L.,  2013, \mn@doi [\mnras]
  {10.1093/mnras/sts595}, \href
  {http://adsabs.harvard.edu/abs/2013MNRAS.429.3353N} {429, 3353}

\bibitem[\protect\citeauthoryear{{Nelson}, {Genel}, {Pillepich},
  {Vogelsberger}, {Springel}  \& {Hernquist}}{{Nelson}
  et~al.}{2016}]{Nelson2016}
{Nelson} D.,  {Genel} S.,  {Pillepich} A.,  {Vogelsberger} M.,  {Springel} V.,
   {Hernquist} L.,  2016, \mn@doi [\mnras] {10.1093/mnras/stw1191}, \href
  {http://adsabs.harvard.edu/abs/2016MNRAS.460.2881N} {460, 2881}

\bibitem[\protect\citeauthoryear{{Pillepich} et~al.,}{{Pillepich}
  et~al.}{2017}]{Pillepich2017}
{Pillepich} A.,  et~al., 2017, preprint, \href
  {http://adsabs.harvard.edu/abs/2017arXiv170302970P} {} (\mn@eprint {arXiv}
  {1703.02970})

\bibitem[\protect\citeauthoryear{{Rahmati}, {Pawlik}, {Rai\v{c}evi\`{c}}  \&
  {Schaye}}{{Rahmati} et~al.}{2013}]{Rahmati2013}
{Rahmati} A.,  {Pawlik} A.~H.,  {Rai\v{c}evi\`{c}} M.,   {Schaye} J.,  2013,
  \mn@doi [\mnras] {10.1093/mnras/stt066}, \href
  {http://adsabs.harvard.edu/abs/2013MNRAS.430.2427R} {430, 2427}

\bibitem[\protect\citeauthoryear{{Saitoh} \& {Makino}}{{Saitoh} \&
  {Makino}}{2010}]{Saitoh2010}
{Saitoh} T.~R.,  {Makino} J.,  2010, \mn@doi [\pasj] {10.1093/pasj/62.2.301},
  \href {http://adsabs.harvard.edu/abs/2010PASJ...62..301S} {62, 301}

\bibitem[\protect\citeauthoryear{{Sijacki}, {Vogelsberger}, {Kere{\v s}},
  {Springel}  \& {Hernquist}}{{Sijacki} et~al.}{2012}]{Sijacki2012}
{Sijacki} D.,  {Vogelsberger} M.,  {Kere{\v s}} D.,  {Springel} V.,
  {Hernquist} L.,  2012, \mn@doi [\mnras] {10.1111/j.1365-2966.2012.21466.x},
  \href {http://adsabs.harvard.edu/abs/2012MNRAS.424.2999S} {424, 2999}

\bibitem[\protect\citeauthoryear{{Smith} et~al.,}{{Smith}
  et~al.}{2017}]{Smith2016}
{Smith} B.~D.,  et~al., 2017, \mn@doi [\mnras] {10.1093/mnras/stw3291}, \href
  {http://adsabs.harvard.edu/abs/2017MNRAS.466.2217S} {466, 2217}

\bibitem[\protect\citeauthoryear{{Springel}}{{Springel}}{2010}]{Arepo}
{Springel} V.,  2010, \mn@doi [\mnras] {10.1111/j.1365-2966.2009.15715.x},
  \href {http://adsabs.harvard.edu/abs/2010MNRAS.401..791S} {401, 791}

\bibitem[\protect\citeauthoryear{{Springel} \& {Hernquist}}{{Springel} \&
  {Hernquist}}{2003}]{Springel2003}
{Springel} V.,  {Hernquist} L.,  2003, \mn@doi [\mnras]
  {10.1046/j.1365-8711.2003.06206.x}, \href
  {http://adsabs.harvard.edu/abs/2003MNRAS.339..289S} {339, 289}

\bibitem[\protect\citeauthoryear{{Springel}, {Yoshida}  \& {White}}{{Springel}
  et~al.}{2001}]{Springel2001}
{Springel} V.,  {Yoshida} N.,   {White} S.~D.~M.,  2001, \mn@doi [\na]
  {10.1016/S1384-1076(01)00042-2}, \href
  {http://adsabs.harvard.edu/abs/2001NewA....6...79S} {6, 79}

\bibitem[\protect\citeauthoryear{{Teyssier}}{{Teyssier}}{2002}]{RAMSES}
{Teyssier} R.,  2002, \mn@doi [\aap] {10.1051/0004-6361:20011817}, \href
  {http://adsabs.harvard.edu/abs/2002A%26A...385..337T} {385, 337}

\bibitem[\protect\citeauthoryear{{Teyssier}}{{Teyssier}}{2015}]{Teyssier2015}
{Teyssier} R.,  2015, \mn@doi [\araa] {10.1146/annurev-astro-082214-122309},
  \href {http://adsabs.harvard.edu/abs/2015ARA%26A..53..325T} {53, 325}

\bibitem[\protect\citeauthoryear{{Torrey}, {Vogelsberger}, {Sijacki},
  {Springel}  \& {Hernquist}}{{Torrey} et~al.}{2012}]{Torrey2012}
{Torrey} P.,  {Vogelsberger} M.,  {Sijacki} D.,  {Springel} V.,   {Hernquist}
  L.,  2012, \mn@doi [\mnras] {10.1111/j.1365-2966.2012.22082.x}, \href
  {http://adsabs.harvard.edu/abs/2012MNRAS.427.2224T} {427, 2224}

\bibitem[\protect\citeauthoryear{{Townsend}}{{Townsend}}{2009}]{Townsend2009}
{Townsend} R.~H.~D.,  2009, \mn@doi [\apjs] {10.1088/0067-0049/181/2/391},
  \href {http://adsabs.harvard.edu/abs/2009ApJS..181..391T} {181, 391}

\bibitem[\protect\citeauthoryear{{Vogelsberger}, {Sijacki}, {Kere{\v s}},
  {Springel}  \& {Hernquist}}{{Vogelsberger} et~al.}{2012}]{Vogelsberger2012}
{Vogelsberger} M.,  {Sijacki} D.,  {Kere{\v s}} D.,  {Springel} V.,
  {Hernquist} L.,  2012, \mn@doi [\mnras] {10.1111/j.1365-2966.2012.21590.x},
  \href {http://adsabs.harvard.edu/abs/2012MNRAS.425.3024V} {425, 3024}

\bibitem[\protect\citeauthoryear{{Vogelsberger}, {Genel}, {Sijacki}, {Torrey},
  {Springel}  \& {Hernquist}}{{Vogelsberger} et~al.}{2013}]{Vogelsberger2013}
{Vogelsberger} M.,  {Genel} S.,  {Sijacki} D.,  {Torrey} P.,  {Springel} V.,
  {Hernquist} L.,  2013, \mn@doi [\mnras] {10.1093/mnras/stt1789}, \href
  {http://adsabs.harvard.edu/abs/2013MNRAS.436.3031V} {436, 3031}

\bibitem[\protect\citeauthoryear{{Weinberger} et~al.,}{{Weinberger}
  et~al.}{2017}]{Weinberger2017}
{Weinberger} R.,  et~al., 2017, \mn@doi [\mnras] {10.1093/mnras/stw2944}, \href
  {http://adsabs.harvard.edu/abs/2017MNRAS.465.3291W} {465, 3291}

\bibitem[\protect\citeauthoryear{{Wiersma}, {Schaye}  \& {Smith}}{{Wiersma}
  et~al.}{2009}]{Wiersma2009}
{Wiersma} R.~P.~C.,  {Schaye} J.,   {Smith} B.~D.,  2009, \mn@doi [\mnras]
  {10.1111/j.1365-2966.2008.14191.x}, \href
  {http://adsabs.harvard.edu/abs/2009MNRAS.393...99W} {393, 99}

\bibitem[\protect\citeauthoryear{{Zhu} \& {Li}}{{Zhu} \& {Li}}{2016}]{Zhu2016}
{Zhu} Q.,  {Li} Y.,  2016, \mn@doi [\apj] {10.3847/0004-637X/831/1/52}, \href
  {http://adsabs.harvard.edu/abs/2016ApJ...831...52Z} {831, 52}

\bibitem[\protect\citeauthoryear{{Zhu}, {Hernquist}  \& {Li}}{{Zhu}
  et~al.}{2015}]{Zhu2015}
{Zhu} Q.,  {Hernquist} L.,   {Li} Y.,  2015, \mn@doi [\apj]
  {10.1088/0004-637X/800/1/6}, \href
  {http://adsabs.harvard.edu/abs/2015ApJ...800....6Z} {800, 6}

\makeatother
\end{thebibliography}

\bsp	
\label{lastpage}
\end{document}